\documentstyle[12pt]{article}
\newtheorem{th}{Theorem}
\newtheorem{lemm}{Lemma}

\def\sar#1{\stackrel{#1}{\longrightarrow}}
\def\slr#1{\stackrel{#1}{\longleftarrow}}
\def\sd{\stackrel{def}{=}}
\def\sq{\stackrel{?}{=}}
\def\0{(0,0,0)}
\def\p#1#2#3#4{P^{(#1,#2)}_{x_{#1}^{#3}x_{#2}^{#4}}}
\def\la{\lambda}
\def\proof{{\sl Proof }}
\def\be{\begin{equation}}
\def\ee#1{\label{#1}\end{equation}}
\def\a{\alpha}
\def\-{\overline }
\def\b{\beta}
\def\B{B\"acklund }
\def\teta{\vartheta}
\def\R{Ribaucour}
\def\ds{\displaystyle}
\def\s{\\[0.5em]}
\def\const{{\rm const}}
\def\tr{transformation}

\begin{document}

\author{Ganzha E.I., Tsarev S.P. \\
  Dept. Mathematics, \\
  Krasnoyarsk State Pedagogical University \\
  Lebedevoi, 89, 660049, Krasnoyarsk, RUSSIA, \\
 e-mail: tsarev@edk.krasnoyarsk.su }

\title{ On superposition of the autoB\"acklund transformations for
$(2+1)$-dimensional integrable systems\thanks{The research described in
this contribution was supported in part by
grant RFBR-DFG No 96-01-00050}}

\date{9 June 1996}
\maketitle

The well known classical auto\B \tr\ for    the sine-Gordon equation
\be
 u_{xt}=\sin u
\ee{SG}
given by (cf. \cite{rog-sha})
\be
\left\{
\begin{array}{l}
\ds
\-u_t = u_t - 2 \a \sin \frac{u+\-u}{2} \s
\ds
\-u_x =-u_x + \frac{2}{ \a} \sin \frac{u-\-u}{2}
\end{array}
\right.
\ee{KV}
(where $\a= \const$ is a parameter of the \tr), possesses the
nonlinear superposition property connecting solutions
 $u^{(0)}$, $u^{({1})}$, $u^{({2})}$ and $u^{({12})}$ of (\ref{SG})
if the pairs $(u^{(0)},u^{({1})})$,
$(u^{({2})}, u^{({12})})$ are connected by (\ref{KV})
with  $\a=\a_1$, and
$(u^{(0)}, u^{({2})})$,
$(u^{({1})},u^{({12})})$  are connected by (\ref{KV})
with $\a=\a_2$:
\be
u^{({12})} = 4 \; \mbox{atan} \left[\frac{\a_1 + \a_2}{\a_1-\a_2} \mbox{tan}
    \left(\frac{u^{({1})}-u^{({2})}}{4}\right)\right] + u^{(0)}.
\ee{KS}
The formula (\ref{KS}) is schematically presented on Fig. 1.

\unitlength=1.00mm
\linethickness{0.4pt}
\begin{picture}(127.00,112.00)
\put(20.00,80.00){\circle{14.00}}
\put(120.00,80.00){\circle{14.00}}
\put(70.00,105.00){\circle{14.00}}
\put(70.00,55.00){\circle{14.00}}
\put(20.00,80.00){\makebox(0,0)[cc]{$u^{(0)}$}}
\put(70.00,105.00){\makebox(0,0)[cc]{$u^{(2)}$}}
\put(70.00,55.00){\makebox(0,0)[cc]{$u^{(1)}$}}
\put(120.00,80.00){\makebox(0,0)[cc]{$u^{(12)}$}}
\put(115.00,85.00){\vector(-3,1){40.00}}
\put(75.00,98.39){\vector(3,-1){40.00}}
\put(115.00,75.00){\vector(-3,-1){40.00}}
\put(75.00,61.60){\vector(3,1){40.00}}
\put(25.00,75.00){\vector(3,-1){40.00}}
\put(65.00,61.60){\vector(-3,1){40.00}}
\put(25.00,85.00){\vector(3,1){40.00}}
\put(65.00,98.39){\vector(-3,-1){40.00}}
\put(40.00,95.00){\makebox(0,0)[cc]{$\a_2$}}  
\put(100.00,95.00){\makebox(0,0)[cc]{$\a_1$}}
\put(40.00,65.00){\makebox(0,0)[cc]{$\a_1$}}
\put(100.00,65.00){\makebox(0,0)[cc]{$\a_2$}}  
\put(70.00,30.00){\makebox(0,0)[cc]{Fig. 1}}
\end{picture}

Since (\ref{KV}) defines the result of the \B \tr\ up to    1 constant
 (the initial data for $\-u$) we need the following precise formulation:
{\sl among the \B \tr s with the    parameter $\a_2$, obtained from
 $u^{(1)}$ and among the \B \tr s with the    parameter
 $\a_1$, obtained from $u^{(2)}$ one can find exactly one common solution
of (\ref{SG}) defined by the explicit formula (\ref{KS}).}
Hereafter we will use the word "\tr " instead of    "auto\tr "

The aforementioned nonlinear superposition principle
("the permutability property") was discovered by L.Bianchi
in the framework of the classical differential geometry when he
studied the induced by the \B \tr s deformations    of constant
curvature surfaces in ${\bf R}^3$. In that period L.Bianchi, G.Darboux
and others  (see \cite{darboux,bianchi,tzitz10}) found
\B \tr s    also for the equation $u_{xt}= e^u - e^{-2u}$,
the "3-wave resonant interaction" system
$$
\left\{
\begin{array}{l}
\ds
u^1_t = c_1u^1_x +\kappa u^2u^3, \s
\ds
u^2_t = c_2u^2_x +\kappa u^1u^3, \s
\ds
u^3_t = c_3u^3_x +\kappa u^1u^2,
\end{array}
\right.
$$
(the latter coincides with the theory
of Egorov type curvilinear coordinate
systems in ${\bf R}^3$) and some others. They also found the
appropriate nonlinear superposition formulas for these systems.
In the  modern theory of $(1+1)$-dimensional
nonlinear integrable systems of partial differential equations
\B \tr s     (see \cite{rog-sha,miura}) and the nonlinear
superposition formulas were found for the majority of all known
integrable systems.

For $(2+1)$-dimensional integrable systems (i.e. systems with 2
"spatial" and one "time" independent variables) some examples of
\B \tr s    were also found. In the classical differential geometry
of XIX--beginning of the XX century the respective problems
(described by partial differential systems with 3 independent variables)
were also intensively studied: the theory of arbitrary
curvilinear orthogonal coordinate systems
in ${\bf R}^3$ (and in ${\bf R}^n$ which is again only
a $(2+1)$-dimensional problem for any $n$), the theory of
conjugate coordinate systems, numerous problems in the theory of
congruences
(\cite{darboux,bianchi,finikov,eis-tr}). For these problems the
proper theory of \B \tr s was constructed alongside with    the
nonlinear superposition principle ("permutability property").
But their properties are more complicated and are given in
Table~1 compared to the properties of \B \tr s of
 $(1+1)$-dimensional systems.

\begin{table}[h]
\begin{tabular}{|c|p{3.5cm}|p{5.5cm}|p{5.5cm}|}
\hline
 &  & $(1+1)$-dim case &  $(2+1)$-dim case
     \\ \hline
 1  & Solutions of the system   & solutions are parameterized by (several) functions of
  1 variable  &   solutions are parameterized by (several) functions of
  2 variables   \\  \hline
2   &  \B \tr\  & Gives a new solution parameterized by (several)
constants (as a solution of ODE's)  &
  Gives a new solution parameterized by (several) functions of 1
variable  \\  \hline
3   & The solution $u^{(12)}$ on Fig. 1   & defined uniquely by
an algebraic formula via $u^{(0)}$, $u^{(1)}$, $u^{(2)}$  &
is parameterized by (several) constants (and usually found via a quadrature),
i.e. we have no algebraic formula \\ \hline
\end{tabular}
\label{tab1}
\end{table}

In the modern theory of $(2+1)$-dimensional integrable systems (see
\cite{oev-rog,ath-nim,kon-sch-rog}) the appropriate \B \tr s also
possess the property 2 of Table 1, and in the cases where a
superposition formula was given the property 3 holds.
As noted in
 \cite{ath-nim,nim1} in the majority of these cases the \B \tr s
were induced by the so called Moutard \tr\ well known
 in the classical differential geometry (see below).

In the present paper we will show how in the case of
$(2+1)$-dimensional integrable systems one can find an extended
formula of nonlinear superposition such that the resulting
solution will be found uniquely from the given previous solution
with algebraic operations.

The exposition of the method will be given for the case of the
so called \R\ \tr s of triply orthogonal    curvilinear coordinate
systems in ${\bf R}^3$ --- one of the
$(2+1)$-dimensional \B \tr s obtained in    classical differential geometry
as well as for the Moutard \tr.

Let us remind the respective definitions (cf. \cite{dar-ort,bianchi}).

{\bf Definition 1.} Two surfaces
 $S_1$ and $S_2$ are called related by a \R\ \tr\    if one can
establish (locally) a one-to-one correspondence between their points
 $(x_1 \in S_1 )\leftrightarrow (x_2 \in S_2)$, such that the
normals to
$S_i$ taken in the corresponding point intersect in a point
 $P=P(x_1)$ such that
 $|Px_1|=|Px_2|$ and the curvature lines of
 $S_1$ correspond to the curvature lines of $S_2$.

Hence  $S_1$ and $S_2$  are two corresponding pieces of the
enveloping surface of some 2-parametric family of spheres
with the additional property of correspondence of curvature
lines.

In order to generalize the \R\ \tr\ for the case    of two triply
orthogonal curvilinear coordinate systems
in   ${\bf R}^3$  we have to take three 3-parametric families of
surfaces (or $n$-parametric for the $n$-dimensional case)
tangent in the corresponding (i.e. having the same values of the
curvilinear coordinates) points to the corresponding
one-parametric families of coordinate surfaces of the both
orthogonal coordinate systems. Since due to the Dupin theorem the
coordinate lines (i.e. lines of intersection of the coordinate
surfaces) of any triply orthogonal curvilinear coordinate system
are the curvature lines on the coordinate surfaces the described
correspondence will guarantee the correspondence of curvature lines.
Let us give the formulas (see \cite{dar-ort,bianchi}). Hereafter one shall
NOT apply summation to repeated indices unless  stated explicitly.

Let  $x^i = x^i(u^1, u^2,u^3)$ be a curvilinear orthogonal
coordinate system. The quantities
  $H_i = \left|\partial_i\vec x \right|$, $\partial_i =\partial
/\partial u^i$, are called its Lam\'e coefficients, and
 $\b_{ik} =\partial_i H_k\big/
H_i$, $i \neq k$, are called its rotation coefficients.
 $\Gamma^k_{ki}= \partial_i H_k
\big/ H_k$ are the Christoffel symbols of the corresponding
diagonal metric
$g_{ii}(u) = H_i^2(u)$ and $\vec X_i = \partial_i \vec x \big/
H_i$ gives the orthonormal tangent frame at the point $\vec x$.
We have the known Lam\'e system of equations (necessary and
sufficient for $\b_{ij}$ to be the rotation coefficients of some
orthogonal curvilinear coordinate system):
\be
\left\{
\begin{array}{l}
\partial_i\b_{jk} = \b_{ji}\b_{ik}, \quad i \neq j \neq k, \s
\partial_i\b_{ik} +\partial_k\b_{ki} + \b_{ji}\b_{jk} = 0,
     \quad i \neq j \neq k.
\end{array}
\right.
\ee{UL}
The system (\ref{UL}) is an overdetermined system whose
solutions are parameterized by 3 functions of 2 variables (see
\cite{dar-ort,bianchi}). In order to define a \R\ \tr\ in
terms of the rotation coefficients we have to find a solution
("a potential") of
the following  overdetermined compatible system
\be
\partial_i\partial_j \varphi = \Gamma^i_{ij}\partial_i\varphi
+ \Gamma^j_{ji}\partial_j \varphi, \quad i \neq j.
\ee{UFI}
Then the \R\ \tr\ will be given by the formulas \cite{bianchi}
\be
\widetilde{\vec x} = \vec x - \frac{2\varphi }{A} \sum_{i=1}^3
\gamma_i \vec X_i,
\ee{PIKS}
\be
\begin{array}{l}
\ds
\widetilde{\vec X_i} = \vec X_i - \frac{2\gamma_i }{A} \sum_{s=1}^3
\gamma_s \vec X_s, \s
\ds
\widetilde{H_i} = H_i - \frac{ 2 \varphi }{A} \Big(\partial_i \gamma_i
+ \sum_{s \neq i} \b_{si}\gamma_s\Big), \s
\ds
\widetilde{\b_{ik}} = \b_{ik} - \frac{2\gamma_i }{A} \Big(\partial_k \gamma_k
+ \sum_{s \neq k} \b_{sk}\gamma_s\Big),
\end{array}
\ee{PVSEH}
where $\gamma_i = \partial_i\varphi \big/ H_i$, $A= \sum (\gamma_i)^2$.
The nonlinear superposition of two \R\ \tr s
 $\vec x^{(0)} \sar{\varphi^{(1)}}
\vec x^{(1)}$, $\vec x^{(0)} \sar{\varphi^{(2)}} \vec x^{(2)}$,
i.e. \tr    s
$\vec x^{(1)} \sar{\varphi^{(12)}} \vec x^{(12)}$,
$\vec x^{(2)} \sar{\varphi^{(21)}} \vec x^{(12)}$,
is given by (\ref{PIKS}), (\ref{PVSEH}) where
\be
\begin{array}{l}
\ds
\varphi^{(12)} =\frac{(a \tau^{(12)} +c)\varphi^{(1)}}{A^{(1)}} +
     a \varphi^{(2)},
\quad
\varphi^{(21)} =- \frac{(a \tau^{(12)} +c+2 a
 B^{(12)})\varphi^{(2)}}{A^{(2)}} +
     a \varphi^{(1)}, \s
\ds
\gamma_i^{(12)} =\frac{(a \tau^{(12)} +c)\gamma_i^{(1)}}{A^{(1)}} +
     a \gamma_i^{(2)},
\quad
\gamma_i^{(21)} = - \frac{(a \tau^{(12)}
+c+2 a B^{(12)})\gamma_i^{(2)}}{A^{(2)}} +     a \gamma_i^{(1)},
\end{array}
\ee{FI}
$A^{(i)}= \sum_s (\gamma_s^{(i)})^2$,
 $B^{(12)} =  \sum_s (\gamma_s^{(1)}\gamma_s^{(2)})$;
$a$ and $c$ are constants and $\tau^{(12)}(u)$ is found via a
quadrature from
\be
\partial_k \tau^{(12)} = -2 \gamma_k^{(2)} \Big(\partial_k\gamma_k^{(1)}+
\sum_{s \neq k} \b_{sk}\gamma_s^{(1)}\Big),
\ee{KVADR}
$\tau^{(21)}=-(\tau^{(12)}+2B^{(12)})$
(line 3 in Table. 1).

The basis of our method is given by the following commutative
diagram of \B \tr s (Fig. 2) where we used the notations for the case of the
\R\ \tr s for the sake of simplicity.

\unitlength 1.00mm
\linethickness{0.4pt}
\begin{picture}(87.00,147.00)
\multiput(30.00,147.00)(0.79,-0.09){2}{\line(1,0){0.79}}
\multiput(31.59,146.82)(0.30,-0.11){5}{\line(1,0){0.30}}
\multiput(33.09,146.28)(0.17,-0.11){8}{\line(1,0){0.17}}
\multiput(34.43,145.42)(0.11,-0.11){10}{\line(0,-1){0.11}}
\multiput(35.54,144.27)(0.12,-0.20){7}{\line(0,-1){0.20}}
\multiput(36.37,142.91)(0.10,-0.30){5}{\line(0,-1){0.30}}
\multiput(36.86,141.39)(0.07,-0.80){2}{\line(0,-1){0.80}}
\multiput(37.00,139.80)(-0.11,-0.79){2}{\line(0,-1){0.79}}
\multiput(36.77,138.22)(-0.12,-0.30){5}{\line(0,-1){0.30}}
\multiput(36.19,136.73)(-0.11,-0.16){8}{\line(0,-1){0.16}}
\multiput(35.29,135.42)(-0.13,-0.12){9}{\line(-1,0){0.13}}
\multiput(34.11,134.34)(-0.20,-0.11){7}{\line(-1,0){0.20}}
\multiput(32.72,133.55)(-0.38,-0.11){4}{\line(-1,0){0.38}}
\put(31.19,133.10){\line(-1,0){1.59}}
\multiput(29.60,133.01)(-0.52,0.09){3}{\line(-1,0){0.52}}
\multiput(28.03,133.28)(-0.25,0.10){6}{\line(-1,0){0.25}}
\multiput(26.56,133.90)(-0.16,0.12){8}{\line(-1,0){0.16}}
\multiput(25.27,134.84)(-0.12,0.13){9}{\line(0,1){0.13}}
\multiput(24.22,136.05)(-0.11,0.20){7}{\line(0,1){0.20}}
\multiput(23.48,137.46)(-0.10,0.39){4}{\line(0,1){0.39}}
\put(23.07,139.00){\line(0,1){1.60}}
\multiput(23.03,140.60)(0.11,0.52){3}{\line(0,1){0.52}}
\multiput(23.34,142.16)(0.11,0.24){6}{\line(0,1){0.24}}
\multiput(24.01,143.61)(0.11,0.14){9}{\line(0,1){0.14}}
\multiput(24.98,144.88)(0.14,0.11){9}{\line(1,0){0.14}}
\multiput(26.22,145.89)(0.24,0.12){6}{\line(1,0){0.24}}
\multiput(27.65,146.59)(0.59,0.10){4}{\line(1,0){0.59}}
\multiput(80.00,147.00)(0.79,-0.09){2}{\line(1,0){0.79}}
\multiput(81.59,146.82)(0.30,-0.11){5}{\line(1,0){0.30}}
\multiput(83.09,146.28)(0.17,-0.11){8}{\line(1,0){0.17}}
\multiput(84.43,145.42)(0.11,-0.11){10}{\line(0,-1){0.11}}
\multiput(85.54,144.27)(0.12,-0.20){7}{\line(0,-1){0.20}}
\multiput(86.37,142.91)(0.10,-0.30){5}{\line(0,-1){0.30}}
\multiput(86.86,141.39)(0.07,-0.80){2}{\line(0,-1){0.80}}
\multiput(87.00,139.80)(-0.11,-0.79){2}{\line(0,-1){0.79}}
\multiput(86.77,138.22)(-0.12,-0.30){5}{\line(0,-1){0.30}}
\multiput(86.19,136.73)(-0.11,-0.16){8}{\line(0,-1){0.16}}
\multiput(85.29,135.42)(-0.13,-0.12){9}{\line(-1,0){0.13}}
\multiput(84.11,134.34)(-0.20,-0.11){7}{\line(-1,0){0.20}}
\multiput(82.72,133.55)(-0.38,-0.11){4}{\line(-1,0){0.38}}
\put(81.19,133.10){\line(-1,0){1.59}}
\multiput(79.60,133.01)(-0.52,0.09){3}{\line(-1,0){0.52}}
\multiput(78.03,133.28)(-0.25,0.10){6}{\line(-1,0){0.25}}
\multiput(76.56,133.90)(-0.16,0.12){8}{\line(-1,0){0.16}}
\multiput(75.27,134.84)(-0.12,0.13){9}{\line(0,1){0.13}}
\multiput(74.22,136.05)(-0.11,0.20){7}{\line(0,1){0.20}}
\multiput(73.48,137.46)(-0.10,0.39){4}{\line(0,1){0.39}}
\put(73.07,139.00){\line(0,1){1.60}}
\multiput(73.03,140.60)(0.11,0.52){3}{\line(0,1){0.52}}
\multiput(73.34,142.16)(0.11,0.24){6}{\line(0,1){0.24}}
\multiput(74.01,143.61)(0.11,0.14){9}{\line(0,1){0.14}}
\multiput(74.98,144.88)(0.14,0.11){9}{\line(1,0){0.14}}
\multiput(76.22,145.89)(0.24,0.12){6}{\line(1,0){0.24}}
\multiput(77.65,146.59)(0.59,0.10){4}{\line(1,0){0.59}}
\multiput(60.00,127.00)(0.79,-0.09){2}{\line(1,0){0.79}}
\multiput(61.59,126.82)(0.30,-0.11){5}{\line(1,0){0.30}}
\multiput(63.09,126.28)(0.17,-0.11){8}{\line(1,0){0.17}}
\multiput(64.43,125.42)(0.11,-0.11){10}{\line(0,-1){0.11}}
\multiput(65.54,124.27)(0.12,-0.20){7}{\line(0,-1){0.20}}
\multiput(66.37,122.91)(0.10,-0.30){5}{\line(0,-1){0.30}}
\multiput(66.86,121.39)(0.07,-0.80){2}{\line(0,-1){0.80}}
\multiput(67.00,119.80)(-0.11,-0.79){2}{\line(0,-1){0.79}}
\multiput(66.77,118.22)(-0.12,-0.30){5}{\line(0,-1){0.30}}
\multiput(66.19,116.73)(-0.11,-0.16){8}{\line(0,-1){0.16}}
\multiput(65.29,115.42)(-0.13,-0.12){9}{\line(-1,0){0.13}}
\multiput(64.11,114.34)(-0.20,-0.11){7}{\line(-1,0){0.20}}
\multiput(62.72,113.55)(-0.38,-0.11){4}{\line(-1,0){0.38}}
\put(61.19,113.10){\line(-1,0){1.59}}
\multiput(59.60,113.01)(-0.52,0.09){3}{\line(-1,0){0.52}}
\multiput(58.03,113.28)(-0.25,0.10){6}{\line(-1,0){0.25}}
\multiput(56.56,113.90)(-0.16,0.12){8}{\line(-1,0){0.16}}
\multiput(55.27,114.84)(-0.12,0.13){9}{\line(0,1){0.13}}
\multiput(54.22,116.05)(-0.11,0.20){7}{\line(0,1){0.20}}
\multiput(53.48,117.46)(-0.10,0.39){4}{\line(0,1){0.39}}
\put(53.07,119.00){\line(0,1){1.60}}
\multiput(53.03,120.60)(0.11,0.52){3}{\line(0,1){0.52}}
\multiput(53.34,122.16)(0.11,0.24){6}{\line(0,1){0.24}}
\multiput(54.01,123.61)(0.11,0.14){9}{\line(0,1){0.14}}
\multiput(54.98,124.88)(0.14,0.11){9}{\line(1,0){0.14}}
\multiput(56.22,125.89)(0.24,0.12){6}{\line(1,0){0.24}}
\multiput(57.65,126.59)(0.59,0.10){4}{\line(1,0){0.59}}
\multiput(10.00,127.00)(0.79,-0.09){2}{\line(1,0){0.79}}
\multiput(11.59,126.82)(0.30,-0.11){5}{\line(1,0){0.30}}
\multiput(13.09,126.28)(0.17,-0.11){8}{\line(1,0){0.17}}
\multiput(14.43,125.42)(0.11,-0.11){10}{\line(0,-1){0.11}}
\multiput(15.54,124.27)(0.12,-0.20){7}{\line(0,-1){0.20}}
\multiput(16.37,122.91)(0.10,-0.30){5}{\line(0,-1){0.30}}
\multiput(16.86,121.39)(0.07,-0.80){2}{\line(0,-1){0.80}}
\multiput(17.00,119.80)(-0.11,-0.79){2}{\line(0,-1){0.79}}
\multiput(16.77,118.22)(-0.12,-0.30){5}{\line(0,-1){0.30}}
\multiput(16.19,116.73)(-0.11,-0.16){8}{\line(0,-1){0.16}}
\multiput(15.29,115.42)(-0.13,-0.12){9}{\line(-1,0){0.13}}
\multiput(14.11,114.34)(-0.20,-0.11){7}{\line(-1,0){0.20}}
\multiput(12.72,113.55)(-0.38,-0.11){4}{\line(-1,0){0.38}}
\put(11.19,113.10){\line(-1,0){1.59}}
\multiput(9.60,113.01)(-0.52,0.09){3}{\line(-1,0){0.52}}
\multiput(8.03,113.28)(-0.25,0.10){6}{\line(-1,0){0.25}}
\multiput(6.56,113.90)(-0.16,0.12){8}{\line(-1,0){0.16}}
\multiput(5.27,114.84)(-0.12,0.13){9}{\line(0,1){0.13}}
\multiput(4.22,116.05)(-0.11,0.20){7}{\line(0,1){0.20}}
\multiput(3.48,117.46)(-0.10,0.39){4}{\line(0,1){0.39}}
\put(3.07,119.00){\line(0,1){1.60}}
\multiput(3.03,120.60)(0.11,0.52){3}{\line(0,1){0.52}}
\multiput(3.34,122.16)(0.11,0.24){6}{\line(0,1){0.24}}
\multiput(4.01,123.61)(0.11,0.14){9}{\line(0,1){0.14}}
\multiput(4.98,124.88)(0.14,0.11){9}{\line(1,0){0.14}}
\multiput(6.22,125.89)(0.24,0.12){6}{\line(1,0){0.24}}
\multiput(7.65,126.59)(0.59,0.10){4}{\line(1,0){0.59}}
\multiput(10.00,77.00)(0.79,-0.09){2}{\line(1,0){0.79}}
\multiput(11.59,76.82)(0.30,-0.11){5}{\line(1,0){0.30}}
\multiput(13.09,76.28)(0.17,-0.11){8}{\line(1,0){0.17}}
\multiput(14.43,75.42)(0.11,-0.11){10}{\line(0,-1){0.11}}
\multiput(15.54,74.27)(0.12,-0.20){7}{\line(0,-1){0.20}}
\multiput(16.37,72.91)(0.10,-0.30){5}{\line(0,-1){0.30}}
\multiput(16.86,71.39)(0.07,-0.80){2}{\line(0,-1){0.80}}
\multiput(17.00,69.80)(-0.11,-0.79){2}{\line(0,-1){0.79}}
\multiput(16.77,68.22)(-0.12,-0.30){5}{\line(0,-1){0.30}}
\multiput(16.19,66.73)(-0.11,-0.16){8}{\line(0,-1){0.16}}
\multiput(15.29,65.42)(-0.13,-0.12){9}{\line(-1,0){0.13}}
\multiput(14.11,64.34)(-0.20,-0.11){7}{\line(-1,0){0.20}}
\multiput(12.72,63.55)(-0.38,-0.11){4}{\line(-1,0){0.38}}
\put(11.19,63.10){\line(-1,0){1.59}}
\multiput(9.60,63.01)(-0.52,0.09){3}{\line(-1,0){0.52}}
\multiput(8.03,63.28)(-0.25,0.10){6}{\line(-1,0){0.25}}
\multiput(6.56,63.90)(-0.16,0.12){8}{\line(-1,0){0.16}}
\multiput(5.27,64.84)(-0.12,0.13){9}{\line(0,1){0.13}}
\multiput(4.22,66.05)(-0.11,0.20){7}{\line(0,1){0.20}}
\multiput(3.48,67.46)(-0.10,0.39){4}{\line(0,1){0.39}}
\put(3.07,69.00){\line(0,1){1.60}}
\multiput(3.03,70.60)(0.11,0.52){3}{\line(0,1){0.52}}
\multiput(3.34,72.16)(0.11,0.24){6}{\line(0,1){0.24}}
\multiput(4.01,73.61)(0.11,0.14){9}{\line(0,1){0.14}}
\multiput(4.98,74.88)(0.14,0.11){9}{\line(1,0){0.14}}
\multiput(6.22,75.89)(0.24,0.12){6}{\line(1,0){0.24}}
\multiput(7.65,76.59)(0.59,0.10){4}{\line(1,0){0.59}}
\multiput(60.00,77.00)(0.79,-0.09){2}{\line(1,0){0.79}}
\multiput(61.59,76.82)(0.30,-0.11){5}{\line(1,0){0.30}}
\multiput(63.09,76.28)(0.17,-0.11){8}{\line(1,0){0.17}}
\multiput(64.43,75.42)(0.11,-0.11){10}{\line(0,-1){0.11}}
\multiput(65.54,74.27)(0.12,-0.20){7}{\line(0,-1){0.20}}
\multiput(66.37,72.91)(0.10,-0.30){5}{\line(0,-1){0.30}}
\multiput(66.86,71.39)(0.07,-0.80){2}{\line(0,-1){0.80}}
\multiput(67.00,69.80)(-0.11,-0.79){2}{\line(0,-1){0.79}}
\multiput(66.77,68.22)(-0.12,-0.30){5}{\line(0,-1){0.30}}
\multiput(66.19,66.73)(-0.11,-0.16){8}{\line(0,-1){0.16}}
\multiput(65.29,65.42)(-0.13,-0.12){9}{\line(-1,0){0.13}}
\multiput(64.11,64.34)(-0.20,-0.11){7}{\line(-1,0){0.20}}
\multiput(62.72,63.55)(-0.38,-0.11){4}{\line(-1,0){0.38}}
\put(61.19,63.10){\line(-1,0){1.59}}
\multiput(59.60,63.01)(-0.52,0.09){3}{\line(-1,0){0.52}}
\multiput(58.03,63.28)(-0.25,0.10){6}{\line(-1,0){0.25}}
\multiput(56.56,63.90)(-0.16,0.12){8}{\line(-1,0){0.16}}
\multiput(55.27,64.84)(-0.12,0.13){9}{\line(0,1){0.13}}
\multiput(54.22,66.05)(-0.11,0.20){7}{\line(0,1){0.20}}
\multiput(53.48,67.46)(-0.10,0.39){4}{\line(0,1){0.39}}
\put(53.07,69.00){\line(0,1){1.60}}
\multiput(53.03,70.60)(0.11,0.52){3}{\line(0,1){0.52}}
\multiput(53.34,72.16)(0.11,0.24){6}{\line(0,1){0.24}}
\multiput(54.01,73.61)(0.11,0.14){9}{\line(0,1){0.14}}
\multiput(54.98,74.88)(0.14,0.11){9}{\line(1,0){0.14}}
\multiput(56.22,75.89)(0.24,0.12){6}{\line(1,0){0.24}}
\multiput(57.65,76.59)(0.59,0.10){4}{\line(1,0){0.59}}
\multiput(30.00,97.00)(0.79,-0.09){2}{\line(1,0){0.79}}
\multiput(31.59,96.82)(0.30,-0.11){5}{\line(1,0){0.30}}
\multiput(33.09,96.28)(0.17,-0.11){8}{\line(1,0){0.17}}
\multiput(34.43,95.42)(0.11,-0.11){10}{\line(0,-1){0.11}}
\multiput(35.54,94.27)(0.12,-0.20){7}{\line(0,-1){0.20}}
\multiput(36.37,92.91)(0.10,-0.30){5}{\line(0,-1){0.30}}
\multiput(36.86,91.39)(0.07,-0.80){2}{\line(0,-1){0.80}}
\multiput(37.00,89.80)(-0.11,-0.79){2}{\line(0,-1){0.79}}
\multiput(36.77,88.22)(-0.12,-0.30){5}{\line(0,-1){0.30}}
\multiput(36.19,86.73)(-0.11,-0.16){8}{\line(0,-1){0.16}}
\multiput(35.29,85.42)(-0.13,-0.12){9}{\line(-1,0){0.13}}
\multiput(34.11,84.34)(-0.20,-0.11){7}{\line(-1,0){0.20}}
\multiput(32.72,83.55)(-0.38,-0.11){4}{\line(-1,0){0.38}}
\put(31.19,83.10){\line(-1,0){1.59}}
\multiput(29.60,83.01)(-0.52,0.09){3}{\line(-1,0){0.52}}
\multiput(28.03,83.28)(-0.25,0.10){6}{\line(-1,0){0.25}}
\multiput(26.56,83.90)(-0.16,0.12){8}{\line(-1,0){0.16}}
\multiput(25.27,84.84)(-0.12,0.13){9}{\line(0,1){0.13}}
\multiput(24.22,86.05)(-0.11,0.20){7}{\line(0,1){0.20}}
\multiput(23.48,87.46)(-0.10,0.39){4}{\line(0,1){0.39}}
\put(23.07,89.00){\line(0,1){1.60}}
\multiput(23.03,90.60)(0.11,0.52){3}{\line(0,1){0.52}}
\multiput(23.34,92.16)(0.11,0.24){6}{\line(0,1){0.24}}
\multiput(24.01,93.61)(0.11,0.14){9}{\line(0,1){0.14}}
\multiput(24.98,94.88)(0.14,0.11){9}{\line(1,0){0.14}}
\multiput(26.22,95.89)(0.24,0.12){6}{\line(1,0){0.24}}
\multiput(27.65,96.59)(0.59,0.10){4}{\line(1,0){0.59}}
\multiput(80.00,97.00)(0.79,-0.09){2}{\line(1,0){0.79}}
\multiput(81.59,96.82)(0.30,-0.11){5}{\line(1,0){0.30}}
\multiput(83.09,96.28)(0.17,-0.11){8}{\line(1,0){0.17}}
\multiput(84.43,95.42)(0.11,-0.11){10}{\line(0,-1){0.11}}
\multiput(85.54,94.27)(0.12,-0.20){7}{\line(0,-1){0.20}}
\multiput(86.37,92.91)(0.10,-0.30){5}{\line(0,-1){0.30}}
\multiput(86.86,91.39)(0.07,-0.80){2}{\line(0,-1){0.80}}
\multiput(87.00,89.80)(-0.11,-0.79){2}{\line(0,-1){0.79}}
\multiput(86.77,88.22)(-0.12,-0.30){5}{\line(0,-1){0.30}}
\multiput(86.19,86.73)(-0.11,-0.16){8}{\line(0,-1){0.16}}
\multiput(85.29,85.42)(-0.13,-0.12){9}{\line(-1,0){0.13}}
\multiput(84.11,84.34)(-0.20,-0.11){7}{\line(-1,0){0.20}}
\multiput(82.72,83.55)(-0.38,-0.11){4}{\line(-1,0){0.38}}
\put(81.19,83.10){\line(-1,0){1.59}}
\multiput(79.60,83.01)(-0.52,0.09){3}{\line(-1,0){0.52}}
\multiput(78.03,83.28)(-0.25,0.10){6}{\line(-1,0){0.25}}
\multiput(76.56,83.90)(-0.16,0.12){8}{\line(-1,0){0.16}}
\multiput(75.27,84.84)(-0.12,0.13){9}{\line(0,1){0.13}}
\multiput(74.22,86.05)(-0.11,0.20){7}{\line(0,1){0.20}}
\multiput(73.48,87.46)(-0.10,0.39){4}{\line(0,1){0.39}}
\put(73.07,89.00){\line(0,1){1.60}}
\multiput(73.03,90.60)(0.11,0.52){3}{\line(0,1){0.52}}
\multiput(73.34,92.16)(0.11,0.24){6}{\line(0,1){0.24}}
\multiput(74.01,93.61)(0.11,0.14){9}{\line(0,1){0.14}}
\multiput(74.98,94.88)(0.14,0.11){9}{\line(1,0){0.14}}
\multiput(76.22,95.89)(0.24,0.12){6}{\line(1,0){0.24}}
\multiput(77.65,96.59)(0.59,0.10){4}{\line(1,0){0.59}}
\put(10.00,70.00){\makebox(0,0)[cc]{$\vec x^{(0)}$}}
\put(60.00,70.00){\makebox(0,0)[cc]{$\vec x^{(1)}$}}
\put(30.00,90.00){\makebox(0,0)[cc]{$\vec x^{(2)}$}}
\put(80.00,90.00){\makebox(0,0)[cc]{$\vec x^{(12)}$}}
\put(10.00,120.00){\makebox(0,0)[cc]{$\vec x^{(3)}$}}
\put(60.00,120.00){\makebox(0,0)[cc]{$\vec x^{(13)}$}}
\put(30.00,140.00){\makebox(0,0)[cc]{$\vec x^{(23)}$}}
\put(80.00,140.00){\makebox(0,0)[cc]{$\vec x^{(123)}$}}
\put(35.00,62.40){\makebox(0,0)[cc]{$\varphi^{(1)}$}}
\put(50.33,96.79){\makebox(0,0)[cc]{$\varphi^{(21)}$}}
\put(75.00,75.00){\makebox(0,0)[cc]{$\varphi^{(12)}$}}
\put(16.67,83.99){\makebox(0,0)[cc]{$\varphi^{(2)}$}}
\put(3.00,95.99){\makebox(0,0)[cc]{$\varphi^{(3)}$}}
\put(85.00,115.00){\makebox(0,0)[cc]{$\varphi^{(123)}$}}
\put(65.33,132.79){\makebox(0,0)[cc]{$\varphi^{(132)}$}}
\put(55.00,145.59){\makebox(0,0)[cc]{$\varphi^{(321)}$}}
\put(10.00,113.59){\line(0,-1){36.80}}
\put(17.00,70.39){\line(1,0){36.00}}
\put(37.00,90.39){\line(1,0){36.00}}
\put(75.00,84.79){\line(-6,-5){10.33}}
\put(60.00,76.79){\line(0,1){36.80}}
\put(80.00,96.79){\line(0,1){36.80}}
\put(30.00,133.59){\line(0,-1){36.80}}
\put(37.33,139.99){\line(1,0){35.67}}
\put(53.00,119.99){\line(-1,0){35.33}}
\put(15.67,124.79){\line(1,1){9.67}}
\put(25.00,85.59){\line(-1,-1){10.67}}
\put(65.33,124.79){\line(1,1){9.67}}
\put(50.00,45.00){\makebox(0,0)[cc]{Fig. 2}}
\put(15.67,134.00){\makebox(0,0)[cc]{$\varphi^{(32)}$}}
\put(41.67,125.67){\makebox(0,0)[cc]{$\varphi^{(31)}$}}
\put(26.33,111.67){\makebox(0,0)[cc]{$\varphi^{(23)}$}}
\put(61.67,100.67){\makebox(0,0)[lc]{$\varphi^{(13)}$}}
\end{picture}

As was shown in  \cite{ts-dis} the diagram on Fig. 2 is actually commutative
for
 $(1+1)$-dimensional equations (sine-Gordon, KdV, the  3-wave system).
In that case if  $u^{(0)}$,
$u^{(1)}$, $u^{(2)}$ are given one can find the solutions
$u^{(12)}$, $u^{(13)}$, $u^{(23)}$ using the algebraic superposition formula
and one had only to check that the last 8th solution
  $u^{(123)}$ will coincide for all three subdiagrams
 $u^{(12)} \rightarrow u^{(123)} \leftarrow u^{(13)}$,
 $u^{(12)} \rightarrow u^{(123)} \leftarrow u^{(23)}$,
 $u^{(13)} \rightarrow u^{(123)} \leftarrow u^{(23)}$ obtainable
again from the algebraic superposition formulas (Fig. 1).
As we show in the following Theorem
 just the diagram on Fig. 2 gives the possibility
to avoid quadratures in the superposition formulas for
 $(2+1)$-dimensional systems and obtain an
algebraic superposition principle ("the \B cube").
\begin{th}
Let three \R\ \tr s
$\vec x^{(0)} \sar{\varphi^{(1)}}
\vec x^{(1)}$,
 $\vec x^{(0)} \sar{\varphi^{(2)}} \vec x^{(2)}$,
 $\vec x^{(0)} \sar{\varphi^{(3)}} \vec x^{(3)}$ of solutions of
 (\ref{UL}) be given as well as obtained from them via
 (\ref{FI}), (\ref{KVADR})
solutions $\vec x^{(12)}$,
$\vec x^{(13)}$, $\vec x^{(23)}$. Then there exists a unique
triply  orthogonal coordinate system
$\vec x^{(123)}$ related by  \R\ \tr s
to $\vec x^{(12)}$, $\vec x^{(13)}$, $\vec x^{(23)}$:
 $\vec x^{(12)} \sar{ \varphi^{(123)}} \vec x^{(123)}$,
 $\vec x^{(13)} \sar{ \varphi^{(132)}} \vec x^{(123)}$,
 $\vec x^{(23)} \sar{ \varphi^{(231)}} \vec x^{(123)}$.
 This solution
 $\vec x^{(123)}$ as well as the corresponding quantities
$H_s^{(123)}$, $\b_{ik}^{(123)}$ may be expressed with algebraic
formulas comprising only the given $\vec x^{(12)}$, $\vec x^{(13)}$, $\vec x^{(23)}$,
$\vec x^{(1)}$, $\vec x^{(2)}$, $\vec x^{(3)}$.
\end{th}
{\sl Proof}. As shown in \cite{bianchi} for the solutions $\vec x^{(0)}(u)$,
$\vec x^{(1)}(u)$, $\vec x^{(2)}(u)$, $\vec x^{(12)}(u)$
connected (see Fig. 1) by \R\ \tr s (\ref{PIKS}), (\ref{PVSEH}),
(\ref{KVADR}) the following holds:
all four points lie on one circumference (depending on the
parameters --- curvilinear coordinates  $u^1$, $u^2$, $u^3$).
Thus if we construct (for some fixed values $u^1$, $u^2$, $u^3$)
the sphere passing through the four points
 $ x^{(0)}$,
$ x^{(1)}$, $ x^{(2)}$, $  x^{(3)}$, then all the other given
$ x^{(12)}$, $ x^{(13)}$, $  x^{(23)}$ will also belong to this sphere.
Draw on this sphere three circumferences passing through the triples
 $( x^{(1)}, x^{(12)}, x^{(13)})$,
 $( x^{(2)}, x^{(12)}, x^{(23)})$,  $( x^{(3)}, x^{(13)},
x^{(23)})$. We will show that all three circumferences will have
one common point --- the point
$ x^{(123)}$ to be found. Indeed after a stereographic
projection of the sphere onto a plane we get the configuration
shown on Fig. 3. (We did not show on this figure only the
circumference passing through
$ x^{(1)}$, $x^{(12)}$, $x^{(13)}$).

\unitlength 1mm
\linethickness{0.4pt}
\begin{picture}(141.93,152.36)
\put(70.00,110.00){\circle*{2.00}}
\put(90.00,110.00){\circle*{2.00}}
\put(90.00,90.00){\circle*{2.00}}
\put(70.00,90.00){\circle*{2.00}}
\put(80.00,114.14){\line(1,0){0.95}}
\put(80.95,114.11){\line(1,0){0.95}}
\multiput(81.90,114.01)(0.47,-0.08){2}{\line(1,0){0.47}}
\multiput(82.84,113.85)(0.46,-0.11){2}{\line(1,0){0.46}}
\multiput(83.77,113.63)(0.30,-0.09){3}{\line(1,0){0.30}}
\multiput(84.68,113.35)(0.30,-0.12){3}{\line(1,0){0.30}}
\multiput(85.56,113.00)(0.22,-0.10){4}{\line(1,0){0.22}}
\multiput(86.43,112.60)(0.21,-0.12){4}{\line(1,0){0.21}}
\multiput(87.26,112.14)(0.16,-0.10){5}{\line(1,0){0.16}}
\multiput(88.06,111.62)(0.15,-0.11){5}{\line(1,0){0.15}}
\multiput(88.83,111.05)(0.12,-0.10){6}{\line(1,0){0.12}}
\multiput(89.55,110.43)(0.11,-0.11){6}{\line(1,0){0.11}}
\multiput(90.23,109.76)(0.11,-0.12){6}{\line(0,-1){0.12}}
\multiput(90.87,109.05)(0.12,-0.15){5}{\line(0,-1){0.15}}
\multiput(91.45,108.30)(0.11,-0.16){5}{\line(0,-1){0.16}}
\multiput(91.98,107.51)(0.12,-0.21){4}{\line(0,-1){0.21}}
\multiput(92.46,106.68)(0.11,-0.21){4}{\line(0,-1){0.21}}
\multiput(92.88,105.83)(0.09,-0.22){4}{\line(0,-1){0.22}}
\multiput(93.25,104.95)(0.10,-0.30){3}{\line(0,-1){0.30}}
\multiput(93.55,104.04)(0.08,-0.31){3}{\line(0,-1){0.31}}
\multiput(93.79,103.12)(0.09,-0.47){2}{\line(0,-1){0.47}}
\put(93.97,102.19){\line(0,-1){0.95}}
\put(94.09,101.24){\line(0,-1){0.95}}
\put(94.14,100.29){\line(0,-1){0.95}}
\put(94.13,99.34){\line(0,-1){0.95}}
\multiput(94.05,98.39)(-0.07,-0.47){2}{\line(0,-1){0.47}}
\multiput(93.91,97.44)(-0.10,-0.47){2}{\line(0,-1){0.47}}
\multiput(93.71,96.51)(-0.09,-0.31){3}{\line(0,-1){0.31}}
\multiput(93.44,95.60)(-0.11,-0.30){3}{\line(0,-1){0.30}}
\multiput(93.11,94.70)(-0.10,-0.22){4}{\line(0,-1){0.22}}
\multiput(92.73,93.83)(-0.11,-0.21){4}{\line(0,-1){0.21}}
\multiput(92.28,92.99)(-0.10,-0.16){5}{\line(0,-1){0.16}}
\multiput(91.78,92.18)(-0.11,-0.16){5}{\line(0,-1){0.16}}
\multiput(91.23,91.40)(-0.10,-0.12){6}{\line(0,-1){0.12}}
\multiput(90.62,90.66)(-0.11,-0.12){6}{\line(0,-1){0.12}}
\multiput(89.97,89.97)(-0.12,-0.11){6}{\line(-1,0){0.12}}
\multiput(89.27,89.32)(-0.12,-0.10){6}{\line(-1,0){0.12}}
\multiput(88.53,88.72)(-0.16,-0.11){5}{\line(-1,0){0.16}}
\multiput(87.75,88.17)(-0.16,-0.10){5}{\line(-1,0){0.16}}
\multiput(86.94,87.68)(-0.21,-0.11){4}{\line(-1,0){0.21}}
\multiput(86.09,87.24)(-0.22,-0.10){4}{\line(-1,0){0.22}}
\multiput(85.22,86.86)(-0.30,-0.11){3}{\line(-1,0){0.30}}
\multiput(84.32,86.53)(-0.31,-0.09){3}{\line(-1,0){0.31}}
\multiput(83.40,86.27)(-0.47,-0.10){2}{\line(-1,0){0.47}}
\multiput(82.47,86.08)(-0.47,-0.07){2}{\line(-1,0){0.47}}
\put(81.53,85.94){\line(-1,0){0.95}}
\put(80.58,85.87){\line(-1,0){0.95}}
\put(79.63,85.86){\line(-1,0){0.95}}
\multiput(78.67,85.92)(-0.47,0.06){2}{\line(-1,0){0.47}}
\multiput(77.73,86.04)(-0.47,0.09){2}{\line(-1,0){0.47}}
\multiput(76.79,86.23)(-0.31,0.08){3}{\line(-1,0){0.31}}
\multiput(75.87,86.47)(-0.30,0.10){3}{\line(-1,0){0.30}}
\multiput(74.97,86.78)(-0.22,0.09){4}{\line(-1,0){0.22}}
\multiput(74.09,87.15)(-0.21,0.11){4}{\line(-1,0){0.21}}
\multiput(73.24,87.58)(-0.16,0.10){5}{\line(-1,0){0.16}}
\multiput(72.42,88.06)(-0.16,0.11){5}{\line(-1,0){0.16}}
\multiput(71.63,88.60)(-0.15,0.12){5}{\line(-1,0){0.15}}
\multiput(70.88,89.19)(-0.12,0.11){6}{\line(-1,0){0.12}}
\multiput(70.18,89.83)(-0.11,0.11){6}{\line(0,1){0.11}}
\multiput(69.51,90.51)(-0.10,0.12){6}{\line(0,1){0.12}}
\multiput(68.90,91.24)(-0.11,0.15){5}{\line(0,1){0.15}}
\multiput(68.33,92.01)(-0.10,0.16){5}{\line(0,1){0.16}}
\multiput(67.82,92.81)(-0.11,0.21){4}{\line(0,1){0.21}}
\multiput(67.36,93.65)(-0.10,0.22){4}{\line(0,1){0.22}}
\multiput(66.97,94.51)(-0.11,0.30){3}{\line(0,1){0.30}}
\multiput(66.63,95.40)(-0.09,0.30){3}{\line(0,1){0.30}}
\multiput(66.35,96.32)(-0.11,0.46){2}{\line(0,1){0.46}}
\multiput(66.13,97.24)(-0.08,0.47){2}{\line(0,1){0.47}}
\put(65.97,98.18){\line(0,1){0.95}}
\put(65.88,99.13){\line(0,1){0.95}}
\put(65.86,100.09){\line(0,1){0.95}}
\put(65.90,101.04){\line(0,1){0.95}}
\multiput(66.00,101.99)(0.08,0.47){2}{\line(0,1){0.47}}
\multiput(66.16,102.92)(0.11,0.46){2}{\line(0,1){0.46}}
\multiput(66.39,103.85)(0.10,0.30){3}{\line(0,1){0.30}}
\multiput(66.68,104.76)(0.12,0.30){3}{\line(0,1){0.30}}
\multiput(67.03,105.64)(0.10,0.22){4}{\line(0,1){0.22}}
\multiput(67.44,106.50)(0.12,0.21){4}{\line(0,1){0.21}}
\multiput(67.91,107.34)(0.10,0.16){5}{\line(0,1){0.16}}
\multiput(68.43,108.13)(0.11,0.15){5}{\line(0,1){0.15}}
\multiput(69.00,108.89)(0.10,0.12){6}{\line(0,1){0.12}}
\multiput(69.63,109.61)(0.11,0.11){6}{\line(0,1){0.11}}
\multiput(70.30,110.29)(0.12,0.10){6}{\line(1,0){0.12}}
\multiput(71.02,110.92)(0.15,0.12){5}{\line(1,0){0.15}}
\multiput(71.77,111.50)(0.16,0.11){5}{\line(1,0){0.16}}
\multiput(72.56,112.03)(0.21,0.12){4}{\line(1,0){0.21}}
\multiput(73.39,112.50)(0.21,0.10){4}{\line(1,0){0.21}}
\multiput(74.25,112.92)(0.29,0.12){3}{\line(1,0){0.29}}
\multiput(75.13,113.28)(0.30,0.10){3}{\line(1,0){0.30}}
\multiput(76.04,113.58)(0.46,0.12){2}{\line(1,0){0.46}}
\multiput(76.96,113.81)(0.47,0.09){2}{\line(1,0){0.47}}
\put(77.90,113.99){\line(1,0){0.95}}
\put(78.84,114.09){\line(1,0){1.16}}
\put(115.00,126.93){\line(1,0){1.50}}
\multiput(116.50,126.88)(0.75,-0.06){2}{\line(1,0){0.75}}
\multiput(118.00,126.76)(0.75,-0.10){2}{\line(1,0){0.75}}
\multiput(119.49,126.55)(0.49,-0.10){3}{\line(1,0){0.49}}
\multiput(120.97,126.26)(0.36,-0.09){4}{\line(1,0){0.36}}
\multiput(122.43,125.88)(0.36,-0.11){4}{\line(1,0){0.36}}
\multiput(123.86,125.43)(0.28,-0.11){5}{\line(1,0){0.28}}
\multiput(125.27,124.89)(0.23,-0.10){6}{\line(1,0){0.23}}
\multiput(126.65,124.28)(0.22,-0.11){6}{\line(1,0){0.22}}
\multiput(127.98,123.59)(0.19,-0.11){7}{\line(1,0){0.19}}
\multiput(129.28,122.83)(0.18,-0.12){7}{\line(1,0){0.18}}
\multiput(130.53,121.99)(0.15,-0.11){8}{\line(1,0){0.15}}
\multiput(131.74,121.09)(0.13,-0.11){9}{\line(1,0){0.13}}
\multiput(132.89,120.12)(0.12,-0.11){9}{\line(1,0){0.12}}
\multiput(133.99,119.09)(0.12,-0.12){9}{\line(0,-1){0.12}}
\multiput(135.02,118.00)(0.11,-0.13){9}{\line(0,-1){0.13}}
\multiput(136.00,116.85)(0.11,-0.15){8}{\line(0,-1){0.15}}
\multiput(136.91,115.65)(0.11,-0.16){8}{\line(0,-1){0.16}}
\multiput(137.75,114.40)(0.11,-0.18){7}{\line(0,-1){0.18}}
\multiput(138.52,113.11)(0.12,-0.22){6}{\line(0,-1){0.22}}
\multiput(139.21,111.78)(0.10,-0.23){6}{\line(0,-1){0.23}}
\multiput(139.83,110.41)(0.11,-0.28){5}{\line(0,-1){0.28}}
\multiput(140.38,109.00)(0.12,-0.36){4}{\line(0,-1){0.36}}
\multiput(140.84,107.57)(0.10,-0.36){4}{\line(0,-1){0.36}}
\multiput(141.22,106.11)(0.10,-0.49){3}{\line(0,-1){0.49}}
\multiput(141.52,104.64)(0.11,-0.74){2}{\line(0,-1){0.74}}
\multiput(141.74,103.15)(0.07,-0.75){2}{\line(0,-1){0.75}}
\put(141.88,101.65){\line(0,-1){1.50}}
\put(141.93,100.15){\line(0,-1){1.50}}
\put(141.89,98.64){\line(0,-1){1.50}}
\multiput(141.77,97.14)(-0.10,-0.75){2}{\line(0,-1){0.75}}
\multiput(141.57,95.65)(-0.09,-0.49){3}{\line(0,-1){0.49}}
\multiput(141.29,94.17)(-0.09,-0.36){4}{\line(0,-1){0.36}}
\multiput(140.92,92.71)(-0.11,-0.36){4}{\line(0,-1){0.36}}
\multiput(140.47,91.27)(-0.11,-0.28){5}{\line(0,-1){0.28}}
\multiput(139.95,89.87)(-0.10,-0.23){6}{\line(0,-1){0.23}}
\multiput(139.34,88.49)(-0.11,-0.22){6}{\line(0,-1){0.22}}
\multiput(138.66,87.14)(-0.11,-0.19){7}{\line(0,-1){0.19}}
\multiput(137.90,85.84)(-0.12,-0.18){7}{\line(0,-1){0.18}}
\multiput(137.08,84.59)(-0.11,-0.15){8}{\line(0,-1){0.15}}
\multiput(136.18,83.38)(-0.11,-0.13){9}{\line(0,-1){0.13}}
\multiput(135.22,82.22)(-0.11,-0.12){9}{\line(0,-1){0.12}}
\multiput(134.19,81.12)(-0.12,-0.12){9}{\line(-1,0){0.12}}
\multiput(133.11,80.07)(-0.13,-0.11){9}{\line(-1,0){0.13}}
\multiput(131.97,79.09)(-0.15,-0.11){8}{\line(-1,0){0.15}}
\multiput(130.77,78.18)(-0.16,-0.11){8}{\line(-1,0){0.16}}
\multiput(129.53,77.33)(-0.18,-0.11){7}{\line(-1,0){0.18}}
\multiput(128.24,76.55)(-0.22,-0.12){6}{\line(-1,0){0.22}}
\multiput(126.91,75.85)(-0.23,-0.10){6}{\line(-1,0){0.23}}
\multiput(125.54,75.22)(-0.28,-0.11){5}{\line(-1,0){0.28}}
\multiput(124.14,74.67)(-0.36,-0.12){4}{\line(-1,0){0.36}}
\multiput(122.71,74.20)(-0.36,-0.10){4}{\line(-1,0){0.36}}
\multiput(121.26,73.81)(-0.49,-0.10){3}{\line(-1,0){0.49}}
\multiput(119.78,73.50)(-0.74,-0.11){2}{\line(-1,0){0.74}}
\multiput(118.29,73.28)(-0.75,-0.07){2}{\line(-1,0){0.75}}
\put(116.80,73.13){\line(-1,0){1.50}}
\put(115.29,73.08){\line(-1,0){1.50}}
\put(113.79,73.10){\line(-1,0){1.50}}
\multiput(112.29,73.21)(-0.75,0.10){2}{\line(-1,0){0.75}}
\multiput(110.79,73.40)(-0.49,0.09){3}{\line(-1,0){0.49}}
\multiput(109.31,73.68)(-0.49,0.12){3}{\line(-1,0){0.49}}
\multiput(107.85,74.04)(-0.36,0.11){4}{\line(-1,0){0.36}}
\multiput(106.41,74.48)(-0.28,0.10){5}{\line(-1,0){0.28}}
\multiput(105.00,75.00)(-0.28,0.12){5}{\line(-1,0){0.28}}
\multiput(103.62,75.60)(-0.22,0.11){6}{\line(-1,0){0.22}}
\multiput(102.27,76.27)(-0.19,0.11){7}{\line(-1,0){0.19}}
\multiput(100.97,77.02)(-0.18,0.12){7}{\line(-1,0){0.18}}
\multiput(99.71,77.84)(-0.15,0.11){8}{\line(-1,0){0.15}}
\multiput(98.49,78.73)(-0.15,0.12){8}{\line(-1,0){0.15}}
\multiput(97.33,79.68)(-0.12,0.11){9}{\line(-1,0){0.12}}
\multiput(96.22,80.70)(-0.12,0.12){9}{\line(0,1){0.12}}
\multiput(95.17,81.78)(-0.11,0.13){9}{\line(0,1){0.13}}
\multiput(94.19,82.92)(-0.12,0.15){8}{\line(0,1){0.15}}
\multiput(93.26,84.11)(-0.11,0.15){8}{\line(0,1){0.15}}
\multiput(92.41,85.35)(-0.11,0.18){7}{\line(0,1){0.18}}
\multiput(91.63,86.63)(-0.12,0.22){6}{\line(0,1){0.22}}
\multiput(90.92,87.96)(-0.11,0.23){6}{\line(0,1){0.23}}
\multiput(90.28,89.33)(-0.11,0.28){5}{\line(0,1){0.28}}
\multiput(89.72,90.72)(-0.12,0.36){4}{\line(0,1){0.36}}
\multiput(89.24,92.15)(-0.10,0.36){4}{\line(0,1){0.36}}
\multiput(88.85,93.60)(-0.11,0.49){3}{\line(0,1){0.49}}
\multiput(88.53,95.07)(-0.12,0.74){2}{\line(0,1){0.74}}
\multiput(88.29,96.56)(-0.08,0.75){2}{\line(0,1){0.75}}
\put(88.14,98.06){\line(0,1){1.50}}
\put(88.08,99.56){\line(0,1){1.51}}
\put(88.10,101.07){\line(0,1){1.50}}
\multiput(88.20,102.57)(0.09,0.75){2}{\line(0,1){0.75}}
\multiput(88.38,104.06)(0.09,0.49){3}{\line(0,1){0.49}}
\multiput(88.65,105.54)(0.12,0.49){3}{\line(0,1){0.49}}
\multiput(89.00,107.01)(0.11,0.36){4}{\line(0,1){0.36}}
\multiput(89.43,108.45)(0.10,0.28){5}{\line(0,1){0.28}}
\multiput(89.95,109.86)(0.12,0.28){5}{\line(0,1){0.28}}
\multiput(90.54,111.25)(0.11,0.22){6}{\line(0,1){0.22}}
\multiput(91.20,112.60)(0.11,0.19){7}{\line(0,1){0.19}}
\multiput(91.94,113.91)(0.12,0.18){7}{\line(0,1){0.18}}
\multiput(92.76,115.17)(0.11,0.15){8}{\line(0,1){0.15}}
\multiput(93.64,116.39)(0.12,0.15){8}{\line(0,1){0.15}}
\multiput(94.59,117.56)(0.11,0.12){9}{\line(0,1){0.12}}
\multiput(95.60,118.67)(0.12,0.12){9}{\line(1,0){0.12}}
\multiput(96.68,119.73)(0.13,0.11){9}{\line(1,0){0.13}}
\multiput(97.81,120.72)(0.15,0.12){8}{\line(1,0){0.15}}
\multiput(98.99,121.65)(0.15,0.11){8}{\line(1,0){0.15}}
\multiput(100.23,122.51)(0.18,0.11){7}{\line(1,0){0.18}}
\multiput(101.51,123.30)(0.22,0.12){6}{\line(1,0){0.22}}
\multiput(102.83,124.02)(0.23,0.11){6}{\line(1,0){0.23}}
\multiput(104.19,124.66)(0.28,0.11){5}{\line(1,0){0.28}}
\multiput(105.59,125.23)(0.28,0.10){5}{\line(1,0){0.28}}
\multiput(107.01,125.71)(0.36,0.10){4}{\line(1,0){0.36}}
\multiput(108.46,126.12)(0.49,0.11){3}{\line(1,0){0.49}}
\multiput(109.93,126.44)(0.50,0.08){3}{\line(1,0){0.50}}
\multiput(111.41,126.69)(0.75,0.08){2}{\line(1,0){0.75}}
\put(112.91,126.84){\line(1,0){2.09}}
\put(50.00,122.36){\line(1,0){1.33}}
\put(51.33,122.32){\line(1,0){1.33}}
\multiput(52.66,122.20)(0.66,-0.10){2}{\line(1,0){0.66}}
\multiput(53.97,122.00)(0.43,-0.09){3}{\line(1,0){0.43}}
\multiput(55.27,121.73)(0.43,-0.12){3}{\line(1,0){0.43}}
\multiput(56.56,121.38)(0.32,-0.11){4}{\line(1,0){0.32}}
\multiput(57.82,120.95)(0.25,-0.10){5}{\line(1,0){0.25}}
\multiput(59.05,120.45)(0.24,-0.11){5}{\line(1,0){0.24}}
\multiput(60.25,119.87)(0.19,-0.11){6}{\line(1,0){0.19}}
\multiput(61.42,119.23)(0.19,-0.12){6}{\line(1,0){0.19}}
\multiput(62.54,118.51)(0.15,-0.11){7}{\line(1,0){0.15}}
\multiput(63.62,117.73)(0.13,-0.11){8}{\line(1,0){0.13}}
\multiput(64.65,116.89)(0.12,-0.11){8}{\line(1,0){0.12}}
\multiput(65.63,115.99)(0.12,-0.12){8}{\line(0,-1){0.12}}
\multiput(66.55,115.03)(0.11,-0.13){8}{\line(0,-1){0.13}}
\multiput(67.42,114.02)(0.11,-0.15){7}{\line(0,-1){0.15}}
\multiput(68.22,112.96)(0.11,-0.16){7}{\line(0,-1){0.16}}
\multiput(68.96,111.85)(0.11,-0.19){6}{\line(0,-1){0.19}}
\multiput(69.63,110.70)(0.10,-0.20){6}{\line(0,-1){0.20}}
\multiput(70.23,109.52)(0.11,-0.24){5}{\line(0,-1){0.24}}
\multiput(70.76,108.30)(0.11,-0.31){4}{\line(0,-1){0.31}}
\multiput(71.22,107.05)(0.10,-0.32){4}{\line(0,-1){0.32}}
\multiput(71.60,105.77)(0.10,-0.43){3}{\line(0,-1){0.43}}
\multiput(71.91,104.48)(0.11,-0.66){2}{\line(0,-1){0.66}}
\multiput(72.14,103.16)(0.07,-0.66){2}{\line(0,-1){0.66}}
\put(72.28,101.84){\line(0,-1){1.33}}
\put(72.35,100.51){\line(0,-1){1.33}}
\put(72.35,99.18){\line(0,-1){1.33}}
\multiput(72.26,97.85)(-0.08,-0.66){2}{\line(0,-1){0.66}}
\multiput(72.09,96.53)(-0.08,-0.44){3}{\line(0,-1){0.44}}
\multiput(71.84,95.23)(-0.11,-0.43){3}{\line(0,-1){0.43}}
\multiput(71.52,93.93)(-0.10,-0.32){4}{\line(0,-1){0.32}}
\multiput(71.12,92.66)(-0.12,-0.31){4}{\line(0,-1){0.31}}
\multiput(70.65,91.42)(-0.11,-0.24){5}{\line(0,-1){0.24}}
\multiput(70.10,90.21)(-0.10,-0.20){6}{\line(0,-1){0.20}}
\multiput(69.48,89.03)(-0.11,-0.19){6}{\line(0,-1){0.19}}
\multiput(68.80,87.89)(-0.11,-0.16){7}{\line(0,-1){0.16}}
\multiput(68.04,86.79)(-0.12,-0.15){7}{\line(0,-1){0.15}}
\multiput(67.22,85.74)(-0.11,-0.12){8}{\line(0,-1){0.12}}
\multiput(66.35,84.74)(-0.12,-0.12){8}{\line(0,-1){0.12}}
\multiput(65.41,83.80)(-0.12,-0.11){8}{\line(-1,0){0.12}}
\multiput(64.42,82.91)(-0.15,-0.12){7}{\line(-1,0){0.15}}
\multiput(63.38,82.08)(-0.16,-0.11){7}{\line(-1,0){0.16}}
\multiput(62.29,81.32)(-0.19,-0.12){6}{\line(-1,0){0.19}}
\multiput(61.15,80.62)(-0.20,-0.10){6}{\line(-1,0){0.20}}
\multiput(59.98,79.99)(-0.24,-0.11){5}{\line(-1,0){0.24}}
\multiput(58.77,79.43)(-0.25,-0.10){5}{\line(-1,0){0.25}}
\multiput(57.53,78.95)(-0.32,-0.10){4}{\line(-1,0){0.32}}
\multiput(56.27,78.54)(-0.43,-0.11){3}{\line(-1,0){0.43}}
\multiput(54.98,78.20)(-0.44,-0.09){3}{\line(-1,0){0.44}}
\multiput(53.67,77.94)(-0.66,-0.09){2}{\line(-1,0){0.66}}
\put(52.35,77.76){\line(-1,0){1.33}}
\put(51.03,77.66){\line(-1,0){1.33}}
\put(49.69,77.64){\line(-1,0){1.33}}
\multiput(48.36,77.70)(-0.66,0.07){2}{\line(-1,0){0.66}}
\multiput(47.04,77.84)(-0.66,0.11){2}{\line(-1,0){0.66}}
\multiput(45.73,78.05)(-0.43,0.10){3}{\line(-1,0){0.43}}
\multiput(44.43,78.34)(-0.32,0.09){4}{\line(-1,0){0.32}}
\multiput(43.15,78.71)(-0.31,0.11){4}{\line(-1,0){0.31}}
\multiput(41.90,79.16)(-0.25,0.10){5}{\line(-1,0){0.25}}
\multiput(40.67,79.68)(-0.24,0.12){5}{\line(-1,0){0.24}}
\multiput(39.48,80.27)(-0.19,0.11){6}{\line(-1,0){0.19}}
\multiput(38.32,80.93)(-0.16,0.10){7}{\line(-1,0){0.16}}
\multiput(37.21,81.66)(-0.15,0.11){7}{\line(-1,0){0.15}}
\multiput(36.14,82.45)(-0.13,0.11){8}{\line(-1,0){0.13}}
\multiput(35.12,83.31)(-0.12,0.11){8}{\line(-1,0){0.12}}
\multiput(34.15,84.22)(-0.11,0.12){8}{\line(0,1){0.12}}
\multiput(33.24,85.19)(-0.11,0.13){8}{\line(0,1){0.13}}
\multiput(32.39,86.22)(-0.11,0.15){7}{\line(0,1){0.15}}
\multiput(31.60,87.29)(-0.10,0.16){7}{\line(0,1){0.16}}
\multiput(30.88,88.41)(-0.11,0.19){6}{\line(0,1){0.19}}
\multiput(30.22,89.56)(-0.12,0.24){5}{\line(0,1){0.24}}
\multiput(29.64,90.76)(-0.10,0.25){5}{\line(0,1){0.25}}
\multiput(29.12,91.99)(-0.11,0.31){4}{\line(0,1){0.31}}
\multiput(28.68,93.24)(-0.09,0.32){4}{\line(0,1){0.32}}
\multiput(28.32,94.52)(-0.10,0.43){3}{\line(0,1){0.43}}
\multiput(28.03,95.82)(-0.10,0.66){2}{\line(0,1){0.66}}
\multiput(27.82,97.14)(-0.07,0.66){2}{\line(0,1){0.66}}
\put(27.69,98.46){\line(0,1){1.33}}
\put(27.64,99.79){\line(0,1){1.33}}
\put(27.67,101.12){\line(0,1){1.33}}
\multiput(27.77,102.45)(0.09,0.66){2}{\line(0,1){0.66}}
\multiput(27.96,103.77)(0.09,0.43){3}{\line(0,1){0.43}}
\multiput(28.22,105.07)(0.11,0.43){3}{\line(0,1){0.43}}
\multiput(28.56,106.36)(0.10,0.32){4}{\line(0,1){0.32}}
\multiput(28.98,107.62)(0.10,0.25){5}{\line(0,1){0.25}}
\multiput(29.47,108.86)(0.11,0.24){5}{\line(0,1){0.24}}
\multiput(30.03,110.07)(0.11,0.20){6}{\line(0,1){0.20}}
\multiput(30.67,111.24)(0.12,0.19){6}{\line(0,1){0.19}}
\multiput(31.37,112.37)(0.11,0.16){7}{\line(0,1){0.16}}
\multiput(32.14,113.45)(0.12,0.15){7}{\line(0,1){0.15}}
\multiput(32.97,114.49)(0.11,0.12){8}{\line(0,1){0.12}}
\multiput(33.86,115.48)(0.12,0.12){8}{\line(1,0){0.12}}
\multiput(34.81,116.41)(0.13,0.11){8}{\line(1,0){0.13}}
\multiput(35.82,117.29)(0.15,0.12){7}{\line(1,0){0.15}}
\multiput(36.87,118.10)(0.16,0.11){7}{\line(1,0){0.16}}
\multiput(37.97,118.85)(0.19,0.11){6}{\line(1,0){0.19}}
\multiput(39.11,119.53)(0.20,0.10){6}{\line(1,0){0.20}}
\multiput(40.30,120.14)(0.24,0.11){5}{\line(1,0){0.24}}
\multiput(41.51,120.69)(0.31,0.12){4}{\line(1,0){0.31}}
\multiput(42.76,121.16)(0.32,0.10){4}{\line(1,0){0.32}}
\multiput(44.03,121.55)(0.43,0.11){3}{\line(1,0){0.43}}
\multiput(45.32,121.87)(0.65,0.12){2}{\line(1,0){0.65}}
\multiput(46.63,122.11)(0.66,0.08){2}{\line(1,0){0.66}}
\put(47.95,122.27){\line(1,0){2.05}}
\put(80.00,152.36){\line(1,0){1.33}}
\put(81.33,152.32){\line(1,0){1.33}}
\multiput(82.66,152.20)(0.66,-0.10){2}{\line(1,0){0.66}}
\multiput(83.97,152.00)(0.43,-0.09){3}{\line(1,0){0.43}}
\multiput(85.27,151.73)(0.43,-0.12){3}{\line(1,0){0.43}}
\multiput(86.56,151.38)(0.32,-0.11){4}{\line(1,0){0.32}}
\multiput(87.82,150.95)(0.25,-0.10){5}{\line(1,0){0.25}}
\multiput(89.05,150.45)(0.24,-0.11){5}{\line(1,0){0.24}}
\multiput(90.25,149.87)(0.19,-0.11){6}{\line(1,0){0.19}}
\multiput(91.42,149.23)(0.19,-0.12){6}{\line(1,0){0.19}}
\multiput(92.54,148.51)(0.15,-0.11){7}{\line(1,0){0.15}}
\multiput(93.62,147.73)(0.13,-0.11){8}{\line(1,0){0.13}}
\multiput(94.65,146.89)(0.12,-0.11){8}{\line(1,0){0.12}}
\multiput(95.63,145.99)(0.12,-0.12){8}{\line(0,-1){0.12}}
\multiput(96.55,145.03)(0.11,-0.13){8}{\line(0,-1){0.13}}
\multiput(97.42,144.02)(0.11,-0.15){7}{\line(0,-1){0.15}}
\multiput(98.22,142.96)(0.11,-0.16){7}{\line(0,-1){0.16}}
\multiput(98.96,141.85)(0.11,-0.19){6}{\line(0,-1){0.19}}
\multiput(99.63,140.70)(0.10,-0.20){6}{\line(0,-1){0.20}}
\multiput(100.23,139.52)(0.11,-0.24){5}{\line(0,-1){0.24}}
\multiput(100.76,138.30)(0.11,-0.31){4}{\line(0,-1){0.31}}
\multiput(101.22,137.05)(0.10,-0.32){4}{\line(0,-1){0.32}}
\multiput(101.60,135.77)(0.10,-0.43){3}{\line(0,-1){0.43}}
\multiput(101.91,134.48)(0.11,-0.66){2}{\line(0,-1){0.66}}
\multiput(102.14,133.16)(0.07,-0.66){2}{\line(0,-1){0.66}}
\put(102.28,131.84){\line(0,-1){1.33}}
\put(102.35,130.51){\line(0,-1){1.33}}
\put(102.35,129.18){\line(0,-1){1.33}}
\multiput(102.26,127.85)(-0.08,-0.66){2}{\line(0,-1){0.66}}
\multiput(102.09,126.53)(-0.08,-0.44){3}{\line(0,-1){0.44}}
\multiput(101.84,125.23)(-0.11,-0.43){3}{\line(0,-1){0.43}}
\multiput(101.52,123.93)(-0.10,-0.32){4}{\line(0,-1){0.32}}
\multiput(101.12,122.66)(-0.12,-0.31){4}{\line(0,-1){0.31}}
\multiput(100.65,121.42)(-0.11,-0.24){5}{\line(0,-1){0.24}}
\multiput(100.10,120.21)(-0.10,-0.20){6}{\line(0,-1){0.20}}
\multiput(99.48,119.03)(-0.11,-0.19){6}{\line(0,-1){0.19}}
\multiput(98.80,117.89)(-0.11,-0.16){7}{\line(0,-1){0.16}}
\multiput(98.04,116.79)(-0.12,-0.15){7}{\line(0,-1){0.15}}
\multiput(97.22,115.74)(-0.11,-0.12){8}{\line(0,-1){0.12}}
\multiput(96.35,114.74)(-0.12,-0.12){8}{\line(0,-1){0.12}}
\multiput(95.41,113.80)(-0.12,-0.11){8}{\line(-1,0){0.12}}
\multiput(94.42,112.91)(-0.15,-0.12){7}{\line(-1,0){0.15}}
\multiput(93.38,112.08)(-0.16,-0.11){7}{\line(-1,0){0.16}}
\multiput(92.29,111.32)(-0.19,-0.12){6}{\line(-1,0){0.19}}
\multiput(91.15,110.62)(-0.20,-0.10){6}{\line(-1,0){0.20}}
\multiput(89.98,109.99)(-0.24,-0.11){5}{\line(-1,0){0.24}}
\multiput(88.77,109.43)(-0.25,-0.10){5}{\line(-1,0){0.25}}
\multiput(87.53,108.95)(-0.32,-0.10){4}{\line(-1,0){0.32}}
\multiput(86.27,108.54)(-0.43,-0.11){3}{\line(-1,0){0.43}}
\multiput(84.98,108.20)(-0.44,-0.09){3}{\line(-1,0){0.44}}
\multiput(83.67,107.94)(-0.66,-0.09){2}{\line(-1,0){0.66}}
\put(82.35,107.76){\line(-1,0){1.33}}
\put(81.03,107.66){\line(-1,0){1.33}}
\put(79.69,107.64){\line(-1,0){1.33}}
\multiput(78.36,107.70)(-0.66,0.07){2}{\line(-1,0){0.66}}
\multiput(77.04,107.84)(-0.66,0.11){2}{\line(-1,0){0.66}}
\multiput(75.73,108.05)(-0.43,0.10){3}{\line(-1,0){0.43}}
\multiput(74.43,108.34)(-0.32,0.09){4}{\line(-1,0){0.32}}
\multiput(73.15,108.71)(-0.31,0.11){4}{\line(-1,0){0.31}}
\multiput(71.90,109.16)(-0.25,0.10){5}{\line(-1,0){0.25}}
\multiput(70.67,109.68)(-0.24,0.12){5}{\line(-1,0){0.24}}
\multiput(69.48,110.27)(-0.19,0.11){6}{\line(-1,0){0.19}}
\multiput(68.32,110.93)(-0.16,0.10){7}{\line(-1,0){0.16}}
\multiput(67.21,111.66)(-0.15,0.11){7}{\line(-1,0){0.15}}
\multiput(66.14,112.45)(-0.13,0.11){8}{\line(-1,0){0.13}}
\multiput(65.12,113.31)(-0.12,0.11){8}{\line(-1,0){0.12}}
\multiput(64.15,114.22)(-0.11,0.12){8}{\line(0,1){0.12}}
\multiput(63.24,115.19)(-0.11,0.13){8}{\line(0,1){0.13}}
\multiput(62.39,116.22)(-0.11,0.15){7}{\line(0,1){0.15}}
\multiput(61.60,117.29)(-0.10,0.16){7}{\line(0,1){0.16}}
\multiput(60.88,118.41)(-0.11,0.19){6}{\line(0,1){0.19}}
\multiput(60.22,119.56)(-0.12,0.24){5}{\line(0,1){0.24}}
\multiput(59.64,120.76)(-0.10,0.25){5}{\line(0,1){0.25}}
\multiput(59.12,121.99)(-0.11,0.31){4}{\line(0,1){0.31}}
\multiput(58.68,123.24)(-0.09,0.32){4}{\line(0,1){0.32}}
\multiput(58.32,124.52)(-0.10,0.43){3}{\line(0,1){0.43}}
\multiput(58.03,125.82)(-0.10,0.66){2}{\line(0,1){0.66}}
\multiput(57.82,127.14)(-0.07,0.66){2}{\line(0,1){0.66}}
\put(57.69,128.46){\line(0,1){1.33}}
\put(57.64,129.79){\line(0,1){1.33}}
\put(57.67,131.12){\line(0,1){1.33}}
\multiput(57.77,132.45)(0.09,0.66){2}{\line(0,1){0.66}}
\multiput(57.96,133.77)(0.09,0.43){3}{\line(0,1){0.43}}
\multiput(58.22,135.07)(0.11,0.43){3}{\line(0,1){0.43}}
\multiput(58.56,136.36)(0.10,0.32){4}{\line(0,1){0.32}}
\multiput(58.98,137.62)(0.10,0.25){5}{\line(0,1){0.25}}
\multiput(59.47,138.86)(0.11,0.24){5}{\line(0,1){0.24}}
\multiput(60.03,140.07)(0.11,0.20){6}{\line(0,1){0.20}}
\multiput(60.67,141.24)(0.12,0.19){6}{\line(0,1){0.19}}
\multiput(61.37,142.37)(0.11,0.16){7}{\line(0,1){0.16}}
\multiput(62.14,143.45)(0.12,0.15){7}{\line(0,1){0.15}}
\multiput(62.97,144.49)(0.11,0.12){8}{\line(0,1){0.12}}
\multiput(63.86,145.48)(0.12,0.12){8}{\line(1,0){0.12}}
\multiput(64.81,146.41)(0.13,0.11){8}{\line(1,0){0.13}}
\multiput(65.82,147.29)(0.15,0.12){7}{\line(1,0){0.15}}
\multiput(66.87,148.10)(0.16,0.11){7}{\line(1,0){0.16}}
\multiput(67.97,148.85)(0.19,0.11){6}{\line(1,0){0.19}}
\multiput(69.11,149.53)(0.20,0.10){6}{\line(1,0){0.20}}
\multiput(70.30,150.14)(0.24,0.11){5}{\line(1,0){0.24}}
\multiput(71.51,150.69)(0.31,0.12){4}{\line(1,0){0.31}}
\multiput(72.76,151.16)(0.32,0.10){4}{\line(1,0){0.32}}
\multiput(74.03,151.55)(0.43,0.11){3}{\line(1,0){0.43}}
\multiput(75.32,151.87)(0.65,0.12){2}{\line(1,0){0.65}}
\multiput(76.63,152.11)(0.66,0.08){2}{\line(1,0){0.66}}
\put(77.95,152.27){\line(1,0){2.05}}
\put(80.00,91.23){\line(1,0){1.93}}
\multiput(81.93,91.19)(0.97,-0.07){2}{\line(1,0){0.97}}
\multiput(83.87,91.05)(0.96,-0.11){2}{\line(1,0){0.96}}
\multiput(85.79,90.82)(0.64,-0.11){3}{\line(1,0){0.64}}
\multiput(87.70,90.51)(0.47,-0.10){4}{\line(1,0){0.47}}
\multiput(89.59,90.10)(0.37,-0.10){5}{\line(1,0){0.37}}
\multiput(91.46,89.61)(0.37,-0.12){5}{\line(1,0){0.37}}
\multiput(93.31,89.02)(0.30,-0.11){6}{\line(1,0){0.30}}
\multiput(95.12,88.36)(0.25,-0.11){7}{\line(1,0){0.25}}
\multiput(96.91,87.61)(0.25,-0.12){7}{\line(1,0){0.25}}
\multiput(98.65,86.77)(0.21,-0.11){8}{\line(1,0){0.21}}
\multiput(100.36,85.86)(0.18,-0.11){9}{\line(1,0){0.18}}
\multiput(102.02,84.86)(0.18,-0.12){9}{\line(1,0){0.18}}
\multiput(103.63,83.79)(0.16,-0.11){10}{\line(1,0){0.16}}
\multiput(105.19,82.64)(0.14,-0.11){11}{\line(1,0){0.14}}
\multiput(106.69,81.42)(0.13,-0.12){11}{\line(1,0){0.13}}
\multiput(108.14,80.14)(0.12,-0.11){12}{\line(1,0){0.12}}
\multiput(109.52,78.78)(0.12,-0.13){11}{\line(0,-1){0.13}}
\multiput(110.84,77.37)(0.11,-0.13){11}{\line(0,-1){0.13}}
\multiput(112.09,75.89)(0.12,-0.15){10}{\line(0,-1){0.15}}
\multiput(113.27,74.36)(0.11,-0.16){10}{\line(0,-1){0.16}}
\multiput(114.38,72.77)(0.11,-0.18){9}{\line(0,-1){0.18}}
\multiput(115.41,71.13)(0.12,-0.21){8}{\line(0,-1){0.21}}
\multiput(116.36,69.44)(0.11,-0.22){8}{\line(0,-1){0.22}}
\multiput(117.23,67.72)(0.11,-0.25){7}{\line(0,-1){0.25}}
\multiput(118.02,65.95)(0.12,-0.30){6}{\line(0,-1){0.30}}
\multiput(118.73,64.15)(0.10,-0.31){6}{\line(0,-1){0.31}}
\multiput(119.35,62.32)(0.11,-0.37){5}{\line(0,-1){0.37}}
\multiput(119.88,60.46)(0.11,-0.47){4}{\line(0,-1){0.47}}
\multiput(120.33,58.57)(0.12,-0.63){3}{\line(0,-1){0.63}}
\multiput(120.69,56.67)(0.09,-0.64){3}{\line(0,-1){0.64}}
\multiput(120.96,54.75)(0.09,-0.96){2}{\line(0,-1){0.96}}
\put(121.13,52.83){\line(0,-1){1.93}}
\put(121.22,50.89){\line(0,-1){1.94}}
\put(121.22,48.96){\line(0,-1){1.93}}
\put(38.86,47.32){\line(0,1){1.93}}
\put(38.78,49.25){\line(0,1){1.94}}
\put(38.79,51.19){\line(0,1){1.93}}
\multiput(38.89,53.12)(0.10,0.96){2}{\line(0,1){0.96}}
\multiput(39.08,55.05)(0.09,0.64){3}{\line(0,1){0.64}}
\multiput(39.36,56.96)(0.09,0.47){4}{\line(0,1){0.47}}
\multiput(39.73,58.86)(0.12,0.47){4}{\line(0,1){0.47}}
\multiput(40.19,60.74)(0.11,0.37){5}{\line(0,1){0.37}}
\multiput(40.74,62.60)(0.11,0.30){6}{\line(0,1){0.30}}
\multiput(41.38,64.43)(0.12,0.30){6}{\line(0,1){0.30}}
\multiput(42.09,66.22)(0.11,0.25){7}{\line(0,1){0.25}}
\multiput(42.90,67.98)(0.11,0.22){8}{\line(0,1){0.22}}
\multiput(43.78,69.71)(0.11,0.19){9}{\line(0,1){0.19}}
\multiput(44.75,71.38)(0.12,0.18){9}{\line(0,1){0.18}}
\multiput(45.79,73.01)(0.11,0.16){10}{\line(0,1){0.16}}
\multiput(46.91,74.59)(0.12,0.15){10}{\line(0,1){0.15}}
\multiput(48.10,76.12)(0.11,0.13){11}{\line(0,1){0.13}}
\multiput(49.36,77.59)(0.11,0.12){12}{\line(0,1){0.12}}
\multiput(50.69,79.00)(0.12,0.11){12}{\line(1,0){0.12}}
\multiput(52.08,80.34)(0.13,0.12){11}{\line(1,0){0.13}}
\multiput(53.53,81.62)(0.14,0.11){11}{\line(1,0){0.14}}
\multiput(55.05,82.82)(0.16,0.11){10}{\line(1,0){0.16}}
\multiput(56.62,83.96)(0.18,0.12){9}{\line(1,0){0.18}}
\multiput(58.23,85.02)(0.19,0.11){9}{\line(1,0){0.19}}
\multiput(59.90,86.00)(0.21,0.11){8}{\line(1,0){0.21}}
\multiput(61.61,86.90)(0.25,0.12){7}{\line(1,0){0.25}}
\multiput(63.37,87.73)(0.26,0.11){7}{\line(1,0){0.26}}
\multiput(65.15,88.47)(0.30,0.11){6}{\line(1,0){0.30}}
\multiput(66.98,89.12)(0.37,0.11){5}{\line(1,0){0.37}}
\multiput(68.83,89.69)(0.37,0.10){5}{\line(1,0){0.37}}
\multiput(70.70,90.17)(0.47,0.10){4}{\line(1,0){0.47}}
\multiput(72.60,90.56)(0.64,0.10){3}{\line(1,0){0.64}}
\multiput(74.51,90.86)(0.96,0.11){2}{\line(1,0){0.96}}
\multiput(76.43,91.08)(1.78,0.08){2}{\line(1,0){1.78}}
\put(114.00,73.00){\circle*{2.00}}
\put(50.00,78.00){\circle*{2.00}}
\put(60.00,120.00){\circle*{2.00}}
\put(101.00,123.00){\circle*{2.00}}
\put(80.00,36.00){\makebox(0,0)[cc]{Fig. 3}}
\put(37.00,64.00){\makebox(0,0)[cc]{$C_1$}}
\put(33.00,121.00){\makebox(0,0)[cc]{$C_2$}}
\put(62.00,151.00){\makebox(0,0)[cc]{$C_3$}}
\put(135.00,125.00){\makebox(0,0)[cc]{$C_4$}}
\put(71.00,115.00){\makebox(0,0)[cc]{$x^{(0)}$}}
\put(95.00,110.00){\makebox(0,0)[cc]{$x^{(2)}$}}
\put(98.00,127.00){\makebox(0,0)[cc]{$x^{(12)}$}}
\put(55.00,117.00){\makebox(0,0)[cc]{$x^{(1)}$}}
\put(71.00,85.00){\makebox(0,0)[cc]{$x^{(3)}$}}
\put(96.00,91.00){\makebox(0,0)[cc]{$x^{(23)}$}}
\put(50.00,73.00){\makebox(0,0)[cc]{$x^{(13)}$}}
\put(111.00,69.00){\makebox(0,0)[cc]{$x^{(123)}$}}
\end{picture}

Now we have to apply the well known geometric theorem (see, e.g.
\cite[\S 10.9.7.2]{berge}): {\sl if for 4 circumferences
 $C_1$, $C_2$, $C_3$,  $C_4$,
 $C_1 \bigcap C_2 = \{P, P'\}$, $C_2 \bigcap C_3 = \{Q, Q'\}$,
$C_3 \bigcap C_4 = \{R, R'\}$, $C_4 \bigcap C_1 = \{S, S'\}$,
the points $P$, $Q$, $R$, $S$ lie on one circumference
(or a straight line) then also
$P'$, $Q'$, $R'$, $S'$ lie on one circumference (or a straight line).}
So
 $ x^{(1)}$, $x^{(12)}$, $x^{(13)}$, $x^{(123)}$ really lie on one
circumference. In the case when the pole of the stereographic projection
 lie on one of the circumferences the corresponding projection
becomes a straight line. Thus the statement about the
uniqueness of  $x^{(123)}$ is proved. Obviously the coordinates
of the point $x^{(123)}$ are algebraically expressible in terms of
the given
$ x^{(1)}$, $ x^{(2)}$, $  x^{(3)}$, $ x^{(12)}$, $ x^{(13)}$, $  x^{(23)}$.

We still have to show that the found
  $x^{(123)}(u)$ as a function of the parameters $u^i$ gives an
orthogonal curvilinear coordinate system related to
 $ x^{(12)}$, $ x^{(13)}$, $  x^{(23)}$ by \R\ \tr s.

In order to prove this we will show that we can choose at least
some set of potentials $\varphi^{(123)}$,
$\varphi^{(132)}$, $\varphi^{(231)}$
 --- solutions of (\ref{UFI}) with corresponding
to  $ x^{(12)}$, $ x^{(13)}$, $  x^{(23)}$
Christoffel symbols, such that substituting them into (\ref{PIKS})
we will get the common vertex $x^{(123)}$:
 $\vec x^{(12)} \sar{\varphi^{(123)}} \vec x^{(123)}
\slr{\varphi^{(132)}} \vec x^{(13)}$,
 $\vec x^{(12)} \sar{\varphi^{(123)}} \vec x^{(123)}
\slr{\varphi^{(231)}} \vec x^{(23)}$,
 $\vec x^{(13)} \sar{\varphi^{(132)}} \vec x^{(123)}
\slr{\varphi^{(231)}} \vec x^{(23)}$. A priori we can find using
(\ref{FI}), (\ref{KVADR}) some potentials
$\varphi^{(123)} ={(\tau^{(123)}\varphi^{(12)})}/{A^{(12)}} +
      \varphi^{(13)}$,
$\varphi^{(213)} ={( \tau^{(213)}\varphi^{(21)})}/{A^{(21)}} +
      \varphi^{(23)}$,
$\varphi^{(132)} ={(\tau^{(132)}\varphi^{(13)})}/{A^{(13)}} +
      \varphi^{(12)}$,
$\varphi^{(312)} ={( \tau^{(312)}\varphi^{(31)})}/{A^{(31)}} +
      \varphi^{(32)}$,
$\varphi^{(231)} ={(\tau^{(231)}\varphi^{(23)})}/{A^{(23)}} +
      \varphi^{(21)}$,
$\varphi^{(321)} ={( \tau^{(321)}\varphi^{(32)})}/{A^{(32)}} +
      \varphi^{(31)}$,
(where we set for simplicity and whithout loss of generality
the constants $a=1$ and $c=0$ in (\ref{KVADR}).)
which will give us {\em three different} final points
 $\vec x^{(12)} \sar{\varphi^{(123)}} \vec x^{(123)}
\slr{\varphi^{(132)}} \vec x^{(13)}$,
 $\vec x^{(12)} \sar{\varphi^{(213)}} \vec x^{(213)}
\slr{\varphi^{(231)}} \vec x^{(23)}$,
 $\vec x^{(13)} \sar{\varphi^{(312)}} \vec x^{(312)}
\slr{\varphi^{(321)}} \vec x^{(23)}$.
Here the functions $\tau^{(ijk)}$ are (defined up to an additive
constant) solutions of the corresponding equations
(\ref{KVADR}):
\be
\partial_s \tau^{(ijk)} = -2 \gamma_s^{(ik)} \Big(\partial_s\gamma_s^{(ij)}+
\sum_{q \neq s} \b^{(i)}_{qs}\gamma_q^{(ij)}\Big) \s
\ee{KVADR3}
\begin{lemm} The quantities
$\ds \tau^{(123)}= \tau^{(23)} +\frac{ \tau^{(13)} \tau^{(21)}}{A^{(1)}}=
 \tau^{(23)} -\frac{ \tau^{(13)} (\tau^{(12)}+2B^{(12)})}{A^{(1)}}$,
$\ds \tau^{(213)}= \tau^{(13)} +\frac{ \tau^{(12)} \tau^{(23)}}{A^{(2)}}$,
$\ds \tau^{(312)}= \tau^{(12)} +\frac{ \tau^{(13)} \tau^{(32)}}{A^{(3)}}$,
$\ds \tau^{(132)}= \tau^{(32)} +\frac{ \tau^{(31)} \tau^{(12)}}{A^{(1)}}$,
$\ds \tau^{(321)}= \tau^{(21)} +\frac{ \tau^{(23)} \tau^{(31)}}{A^{(3)}}$,
$\ds \tau^{(231)}= \tau^{(31)} +\frac{ \tau^{(32)} \tau^{(21)}}{A^{(2)}}$
satisfy (\ref{KVADR3}) and after substitution into the formulas
for $\varphi^{(ijk)}$ result in
$\varphi^{(123)}=\varphi^{(213)}$, $\varphi^{(132)}=\varphi^{(312)}$,
$\varphi^{(321)}=\varphi^{(231)}$.
\end{lemm}
The proof is easily obtained with a direct computation.

Consequently choosing the given in the Lemma $\tau^{(ijk)}$ we find
ONE common resulting $\vec x^{(123)}=\vec x^{(213)}=\vec x^{(312)}$.
The uniqueness of such $\vec x^{(123)}$ has been proved above.
The proof is completed.

The exposed above method of construction of algebraic nonlinear
superposition formulas is especially simple for the case of the
Moutard \tr\
(see \cite{darboux}, \cite{bianchi}).
This \tr\ played an important role in classical differential
geometry in the theory of
surface deformations, the theory of congruences and nets. We
will formulate the classical definitions in the form convenient
for us (see also the formulation of the \R\ \tr\    in  similar form
in \cite{ganzha2}). Let a "potential" $M(x,y)$ (an arbitrary
function of two variables) is given (for the case of \R\ \tr s
the "potential" $M(u^1, u^2, u^3)$ satisfies a 3rd order equation,
see \cite{dar-ort,ganzha2}). Then a Moutard \tr\    is defined by a solution
  $u= R(x,y)$ of the Moutard equation
\be
u_{xy} = M(x,y)\,u, \qquad u=u(x,y),
\ee{MU1}
with the formulas
\be
M = M_0 \rightarrow M=M_1 =  M_0 - 2(\ln R)_{xy} = -M_0 + \frac{2R_xR_y}{R^2}
  =R\left(\frac1R\right)_{xy}.
\ee{pp}
If we have alongside with  $u(x,y)$ a second solution
$u=\varphi(x,y)$ of the same equation (\ref{MU1})
with the potential $M=M_0$ then one can find using a quadrature
a solution
$\teta$ of the equation (\ref{MU1})
with a transformed potential $M=M_1$:
\be
\left\{\begin{array}{rcl}
\ds
\left(R\teta\right)_x & = & - R^2 \left(\frac{\varphi}{R}\right)_x, \s
\ds
\left(R\teta\right)_y & = &  R^2 \left(\frac{\varphi}{R}\right)_y ,
\end{array}
\right.
\ee{PRM}
Let us interpret
(\ref{PRM}) as a diagram on Fig. 1 (cf. \cite[v. II, p. II, \S
297]{bianchi}): if a potential $M_0$ is given as well as its two
Moutard \tr s
$M_0 \sar{u=R} M_1$, $M_0 \sar{u=\varphi} M_2$ then using a quadrature
we can find
 $M_{12}= M_1 -2 (\log \teta)_{xy}$, i.e. $M_1\sar{\teta}
M_{12}$,  $M_2\sar{\psi} M_{12}$, $\teta$ is found from (\ref{PRM}), $\psi =
-u\teta \big/ \varphi$ (see Fig. 4). L.Bianchi in \cite[v. II, p. II, \S
297]{bianchi} gives in another form the "B\"acklund cube"
formula for the case of Moutard \tr s as    the following proposition:
 {\sl if $R_1$, $R_2$, $\teta$ are three solution of
 (\ref{MU1}) with a given $M=M_0$, and
$\teta_1$, $\teta_2$ (solutions of the transformed equations
with the potentials
$M_1=  M_0 - 2(\ln R_1)_{xy}$, $M_2=  M_0 - 2(\ln R_2)_{xy}$)
are  obtained from
$\teta$ via (\ref{PRM}) then there exists a unique
solution  $\teta'$ of the 4th Moutard equation
 $u_{xy} = M_{12}u$ connected to $\teta_1$,
$\teta_2$ with Moutard \tr s (\ref{PRM}). This $\teta'$ is
expressible with an algebraic formula
$$
\teta' - \teta= \frac{R_1R_2}{\lambda}(\teta_2 - \teta_1),
\qquad \lambda = R_1 R_1'= - R_2 R_2',
$$
where $R_1'$, $R_2'$ are obtained from $R_2$, $R_1$ according to (\ref{PRM}).
}

\unitlength 1.00mm
\linethickness{0.4pt}
\begin{picture}(147.00,147.00)
\multiput(90.00,147.00)(0.79,-0.09){2}{\line(1,0){0.79}}
\multiput(91.59,146.82)(0.30,-0.11){5}{\line(1,0){0.30}}
\multiput(93.09,146.28)(0.17,-0.11){8}{\line(1,0){0.17}}
\multiput(94.43,145.42)(0.11,-0.11){10}{\line(0,-1){0.11}}
\multiput(95.54,144.27)(0.12,-0.20){7}{\line(0,-1){0.20}}
\multiput(96.37,142.91)(0.10,-0.30){5}{\line(0,-1){0.30}}
\multiput(96.86,141.39)(0.07,-0.80){2}{\line(0,-1){0.80}}
\multiput(97.00,139.80)(-0.11,-0.79){2}{\line(0,-1){0.79}}
\multiput(96.77,138.22)(-0.12,-0.30){5}{\line(0,-1){0.30}}
\multiput(96.19,136.73)(-0.11,-0.16){8}{\line(0,-1){0.16}}
\multiput(95.29,135.42)(-0.13,-0.12){9}{\line(-1,0){0.13}}
\multiput(94.11,134.34)(-0.20,-0.11){7}{\line(-1,0){0.20}}
\multiput(92.72,133.55)(-0.38,-0.11){4}{\line(-1,0){0.38}}
\put(91.19,133.10){\line(-1,0){1.59}}
\multiput(89.60,133.01)(-0.52,0.09){3}{\line(-1,0){0.52}}
\multiput(88.03,133.28)(-0.25,0.10){6}{\line(-1,0){0.25}}
\multiput(86.56,133.90)(-0.16,0.12){8}{\line(-1,0){0.16}}
\multiput(85.27,134.84)(-0.12,0.13){9}{\line(0,1){0.13}}
\multiput(84.22,136.05)(-0.11,0.20){7}{\line(0,1){0.20}}
\multiput(83.48,137.46)(-0.10,0.39){4}{\line(0,1){0.39}}
\put(83.07,139.00){\line(0,1){1.60}}
\multiput(83.03,140.60)(0.11,0.52){3}{\line(0,1){0.52}}
\multiput(83.34,142.16)(0.11,0.24){6}{\line(0,1){0.24}}
\multiput(84.01,143.61)(0.11,0.14){9}{\line(0,1){0.14}}
\multiput(84.98,144.88)(0.14,0.11){9}{\line(1,0){0.14}}
\multiput(86.22,145.89)(0.24,0.12){6}{\line(1,0){0.24}}
\multiput(87.65,146.59)(0.59,0.10){4}{\line(1,0){0.59}}
\multiput(140.00,147.00)(0.79,-0.09){2}{\line(1,0){0.79}}
\multiput(141.59,146.82)(0.30,-0.11){5}{\line(1,0){0.30}}
\multiput(143.09,146.28)(0.17,-0.11){8}{\line(1,0){0.17}}
\multiput(144.43,145.42)(0.11,-0.11){10}{\line(0,-1){0.11}}
\multiput(145.54,144.27)(0.12,-0.20){7}{\line(0,-1){0.20}}
\multiput(146.37,142.91)(0.10,-0.30){5}{\line(0,-1){0.30}}
\multiput(146.86,141.39)(0.07,-0.80){2}{\line(0,-1){0.80}}
\multiput(147.00,139.80)(-0.11,-0.79){2}{\line(0,-1){0.79}}
\multiput(146.77,138.22)(-0.12,-0.30){5}{\line(0,-1){0.30}}
\multiput(146.19,136.73)(-0.11,-0.16){8}{\line(0,-1){0.16}}
\multiput(145.29,135.42)(-0.13,-0.12){9}{\line(-1,0){0.13}}
\multiput(144.11,134.34)(-0.20,-0.11){7}{\line(-1,0){0.20}}
\multiput(142.72,133.55)(-0.38,-0.11){4}{\line(-1,0){0.38}}
\put(141.19,133.10){\line(-1,0){1.59}}
\multiput(139.60,133.01)(-0.52,0.09){3}{\line(-1,0){0.52}}
\multiput(138.03,133.28)(-0.25,0.10){6}{\line(-1,0){0.25}}
\multiput(136.56,133.90)(-0.16,0.12){8}{\line(-1,0){0.16}}
\multiput(135.27,134.84)(-0.12,0.13){9}{\line(0,1){0.13}}
\multiput(134.22,136.05)(-0.11,0.20){7}{\line(0,1){0.20}}
\multiput(133.48,137.46)(-0.10,0.39){4}{\line(0,1){0.39}}
\put(133.07,139.00){\line(0,1){1.60}}
\multiput(133.03,140.60)(0.11,0.52){3}{\line(0,1){0.52}}
\multiput(133.34,142.16)(0.11,0.24){6}{\line(0,1){0.24}}
\multiput(134.01,143.61)(0.11,0.14){9}{\line(0,1){0.14}}
\multiput(134.98,144.88)(0.14,0.11){9}{\line(1,0){0.14}}
\multiput(136.22,145.89)(0.24,0.12){6}{\line(1,0){0.24}}
\multiput(137.65,146.59)(0.59,0.10){4}{\line(1,0){0.59}}
\multiput(120.00,127.00)(0.79,-0.09){2}{\line(1,0){0.79}}
\multiput(121.59,126.82)(0.30,-0.11){5}{\line(1,0){0.30}}
\multiput(123.09,126.28)(0.17,-0.11){8}{\line(1,0){0.17}}
\multiput(124.43,125.42)(0.11,-0.11){10}{\line(0,-1){0.11}}
\multiput(125.54,124.27)(0.12,-0.20){7}{\line(0,-1){0.20}}
\multiput(126.37,122.91)(0.10,-0.30){5}{\line(0,-1){0.30}}
\multiput(126.86,121.39)(0.07,-0.80){2}{\line(0,-1){0.80}}
\multiput(127.00,119.80)(-0.11,-0.79){2}{\line(0,-1){0.79}}
\multiput(126.77,118.22)(-0.12,-0.30){5}{\line(0,-1){0.30}}
\multiput(126.19,116.73)(-0.11,-0.16){8}{\line(0,-1){0.16}}
\multiput(125.29,115.42)(-0.13,-0.12){9}{\line(-1,0){0.13}}
\multiput(124.11,114.34)(-0.20,-0.11){7}{\line(-1,0){0.20}}
\multiput(122.72,113.55)(-0.38,-0.11){4}{\line(-1,0){0.38}}
\put(121.19,113.10){\line(-1,0){1.59}}
\multiput(119.60,113.01)(-0.52,0.09){3}{\line(-1,0){0.52}}
\multiput(118.03,113.28)(-0.25,0.10){6}{\line(-1,0){0.25}}
\multiput(116.56,113.90)(-0.16,0.12){8}{\line(-1,0){0.16}}
\multiput(115.27,114.84)(-0.12,0.13){9}{\line(0,1){0.13}}
\multiput(114.22,116.05)(-0.11,0.20){7}{\line(0,1){0.20}}
\multiput(113.48,117.46)(-0.10,0.39){4}{\line(0,1){0.39}}
\put(113.07,119.00){\line(0,1){1.60}}
\multiput(113.03,120.60)(0.11,0.52){3}{\line(0,1){0.52}}
\multiput(113.34,122.16)(0.11,0.24){6}{\line(0,1){0.24}}
\multiput(114.01,123.61)(0.11,0.14){9}{\line(0,1){0.14}}
\multiput(114.98,124.88)(0.14,0.11){9}{\line(1,0){0.14}}
\multiput(116.22,125.89)(0.24,0.12){6}{\line(1,0){0.24}}
\multiput(117.65,126.59)(0.59,0.10){4}{\line(1,0){0.59}}
\multiput(70.00,127.00)(0.79,-0.09){2}{\line(1,0){0.79}}
\multiput(71.59,126.82)(0.30,-0.11){5}{\line(1,0){0.30}}
\multiput(73.09,126.28)(0.17,-0.11){8}{\line(1,0){0.17}}
\multiput(74.43,125.42)(0.11,-0.11){10}{\line(0,-1){0.11}}
\multiput(75.54,124.27)(0.12,-0.20){7}{\line(0,-1){0.20}}
\multiput(76.37,122.91)(0.10,-0.30){5}{\line(0,-1){0.30}}
\multiput(76.86,121.39)(0.07,-0.80){2}{\line(0,-1){0.80}}
\multiput(77.00,119.80)(-0.11,-0.79){2}{\line(0,-1){0.79}}
\multiput(76.77,118.22)(-0.12,-0.30){5}{\line(0,-1){0.30}}
\multiput(76.19,116.73)(-0.11,-0.16){8}{\line(0,-1){0.16}}
\multiput(75.29,115.42)(-0.13,-0.12){9}{\line(-1,0){0.13}}
\multiput(74.11,114.34)(-0.20,-0.11){7}{\line(-1,0){0.20}}
\multiput(72.72,113.55)(-0.38,-0.11){4}{\line(-1,0){0.38}}
\put(71.19,113.10){\line(-1,0){1.59}}
\multiput(69.60,113.01)(-0.52,0.09){3}{\line(-1,0){0.52}}
\multiput(68.03,113.28)(-0.25,0.10){6}{\line(-1,0){0.25}}
\multiput(66.56,113.90)(-0.16,0.12){8}{\line(-1,0){0.16}}
\multiput(65.27,114.84)(-0.12,0.13){9}{\line(0,1){0.13}}
\multiput(64.22,116.05)(-0.11,0.20){7}{\line(0,1){0.20}}
\multiput(63.48,117.46)(-0.10,0.39){4}{\line(0,1){0.39}}
\put(63.07,119.00){\line(0,1){1.60}}
\multiput(63.03,120.60)(0.11,0.52){3}{\line(0,1){0.52}}
\multiput(63.34,122.16)(0.11,0.24){6}{\line(0,1){0.24}}
\multiput(64.01,123.61)(0.11,0.14){9}{\line(0,1){0.14}}
\multiput(64.98,124.88)(0.14,0.11){9}{\line(1,0){0.14}}
\multiput(66.22,125.89)(0.24,0.12){6}{\line(1,0){0.24}}
\multiput(67.65,126.59)(0.59,0.10){4}{\line(1,0){0.59}}
\multiput(70.00,77.00)(0.79,-0.09){2}{\line(1,0){0.79}}
\multiput(71.59,76.82)(0.30,-0.11){5}{\line(1,0){0.30}}
\multiput(73.09,76.28)(0.17,-0.11){8}{\line(1,0){0.17}}
\multiput(74.43,75.42)(0.11,-0.11){10}{\line(0,-1){0.11}}
\multiput(75.54,74.27)(0.12,-0.20){7}{\line(0,-1){0.20}}
\multiput(76.37,72.91)(0.10,-0.30){5}{\line(0,-1){0.30}}
\multiput(76.86,71.39)(0.07,-0.80){2}{\line(0,-1){0.80}}
\multiput(77.00,69.80)(-0.11,-0.79){2}{\line(0,-1){0.79}}
\multiput(76.77,68.22)(-0.12,-0.30){5}{\line(0,-1){0.30}}
\multiput(76.19,66.73)(-0.11,-0.16){8}{\line(0,-1){0.16}}
\multiput(75.29,65.42)(-0.13,-0.12){9}{\line(-1,0){0.13}}
\multiput(74.11,64.34)(-0.20,-0.11){7}{\line(-1,0){0.20}}
\multiput(72.72,63.55)(-0.38,-0.11){4}{\line(-1,0){0.38}}
\put(71.19,63.10){\line(-1,0){1.59}}
\multiput(69.60,63.01)(-0.52,0.09){3}{\line(-1,0){0.52}}
\multiput(68.03,63.28)(-0.25,0.10){6}{\line(-1,0){0.25}}
\multiput(66.56,63.90)(-0.16,0.12){8}{\line(-1,0){0.16}}
\multiput(65.27,64.84)(-0.12,0.13){9}{\line(0,1){0.13}}
\multiput(64.22,66.05)(-0.11,0.20){7}{\line(0,1){0.20}}
\multiput(63.48,67.46)(-0.10,0.39){4}{\line(0,1){0.39}}
\put(63.07,69.00){\line(0,1){1.60}}
\multiput(63.03,70.60)(0.11,0.52){3}{\line(0,1){0.52}}
\multiput(63.34,72.16)(0.11,0.24){6}{\line(0,1){0.24}}
\multiput(64.01,73.61)(0.11,0.14){9}{\line(0,1){0.14}}
\multiput(64.98,74.88)(0.14,0.11){9}{\line(1,0){0.14}}
\multiput(66.22,75.89)(0.24,0.12){6}{\line(1,0){0.24}}
\multiput(67.65,76.59)(0.59,0.10){4}{\line(1,0){0.59}}
\multiput(120.00,77.00)(0.79,-0.09){2}{\line(1,0){0.79}}
\multiput(121.59,76.82)(0.30,-0.11){5}{\line(1,0){0.30}}
\multiput(123.09,76.28)(0.17,-0.11){8}{\line(1,0){0.17}}
\multiput(124.43,75.42)(0.11,-0.11){10}{\line(0,-1){0.11}}
\multiput(125.54,74.27)(0.12,-0.20){7}{\line(0,-1){0.20}}
\multiput(126.37,72.91)(0.10,-0.30){5}{\line(0,-1){0.30}}
\multiput(126.86,71.39)(0.07,-0.80){2}{\line(0,-1){0.80}}
\multiput(127.00,69.80)(-0.11,-0.79){2}{\line(0,-1){0.79}}
\multiput(126.77,68.22)(-0.12,-0.30){5}{\line(0,-1){0.30}}
\multiput(126.19,66.73)(-0.11,-0.16){8}{\line(0,-1){0.16}}
\multiput(125.29,65.42)(-0.13,-0.12){9}{\line(-1,0){0.13}}
\multiput(124.11,64.34)(-0.20,-0.11){7}{\line(-1,0){0.20}}
\multiput(122.72,63.55)(-0.38,-0.11){4}{\line(-1,0){0.38}}
\put(121.19,63.10){\line(-1,0){1.59}}
\multiput(119.60,63.01)(-0.52,0.09){3}{\line(-1,0){0.52}}
\multiput(118.03,63.28)(-0.25,0.10){6}{\line(-1,0){0.25}}
\multiput(116.56,63.90)(-0.16,0.12){8}{\line(-1,0){0.16}}
\multiput(115.27,64.84)(-0.12,0.13){9}{\line(0,1){0.13}}
\multiput(114.22,66.05)(-0.11,0.20){7}{\line(0,1){0.20}}
\multiput(113.48,67.46)(-0.10,0.39){4}{\line(0,1){0.39}}
\put(113.07,69.00){\line(0,1){1.60}}
\multiput(113.03,70.60)(0.11,0.52){3}{\line(0,1){0.52}}
\multiput(113.34,72.16)(0.11,0.24){6}{\line(0,1){0.24}}
\multiput(114.01,73.61)(0.11,0.14){9}{\line(0,1){0.14}}
\multiput(114.98,74.88)(0.14,0.11){9}{\line(1,0){0.14}}
\multiput(116.22,75.89)(0.24,0.12){6}{\line(1,0){0.24}}
\multiput(117.65,76.59)(0.59,0.10){4}{\line(1,0){0.59}}
\multiput(90.00,97.00)(0.79,-0.09){2}{\line(1,0){0.79}}
\multiput(91.59,96.82)(0.30,-0.11){5}{\line(1,0){0.30}}
\multiput(93.09,96.28)(0.17,-0.11){8}{\line(1,0){0.17}}
\multiput(94.43,95.42)(0.11,-0.11){10}{\line(0,-1){0.11}}
\multiput(95.54,94.27)(0.12,-0.20){7}{\line(0,-1){0.20}}
\multiput(96.37,92.91)(0.10,-0.30){5}{\line(0,-1){0.30}}
\multiput(96.86,91.39)(0.07,-0.80){2}{\line(0,-1){0.80}}
\multiput(97.00,89.80)(-0.11,-0.79){2}{\line(0,-1){0.79}}
\multiput(96.77,88.22)(-0.12,-0.30){5}{\line(0,-1){0.30}}
\multiput(96.19,86.73)(-0.11,-0.16){8}{\line(0,-1){0.16}}
\multiput(95.29,85.42)(-0.13,-0.12){9}{\line(-1,0){0.13}}
\multiput(94.11,84.34)(-0.20,-0.11){7}{\line(-1,0){0.20}}
\multiput(92.72,83.55)(-0.38,-0.11){4}{\line(-1,0){0.38}}
\put(91.19,83.10){\line(-1,0){1.59}}
\multiput(89.60,83.01)(-0.52,0.09){3}{\line(-1,0){0.52}}
\multiput(88.03,83.28)(-0.25,0.10){6}{\line(-1,0){0.25}}
\multiput(86.56,83.90)(-0.16,0.12){8}{\line(-1,0){0.16}}
\multiput(85.27,84.84)(-0.12,0.13){9}{\line(0,1){0.13}}
\multiput(84.22,86.05)(-0.11,0.20){7}{\line(0,1){0.20}}
\multiput(83.48,87.46)(-0.10,0.39){4}{\line(0,1){0.39}}
\put(83.07,89.00){\line(0,1){1.60}}
\multiput(83.03,90.60)(0.11,0.52){3}{\line(0,1){0.52}}
\multiput(83.34,92.16)(0.11,0.24){6}{\line(0,1){0.24}}
\multiput(84.01,93.61)(0.11,0.14){9}{\line(0,1){0.14}}
\multiput(84.98,94.88)(0.14,0.11){9}{\line(1,0){0.14}}
\multiput(86.22,95.89)(0.24,0.12){6}{\line(1,0){0.24}}
\multiput(87.65,96.59)(0.59,0.10){4}{\line(1,0){0.59}}
\multiput(140.00,97.00)(0.79,-0.09){2}{\line(1,0){0.79}}
\multiput(141.59,96.82)(0.30,-0.11){5}{\line(1,0){0.30}}
\multiput(143.09,96.28)(0.17,-0.11){8}{\line(1,0){0.17}}
\multiput(144.43,95.42)(0.11,-0.11){10}{\line(0,-1){0.11}}
\multiput(145.54,94.27)(0.12,-0.20){7}{\line(0,-1){0.20}}
\multiput(146.37,92.91)(0.10,-0.30){5}{\line(0,-1){0.30}}
\multiput(146.86,91.39)(0.07,-0.80){2}{\line(0,-1){0.80}}
\multiput(147.00,89.80)(-0.11,-0.79){2}{\line(0,-1){0.79}}
\multiput(146.77,88.22)(-0.12,-0.30){5}{\line(0,-1){0.30}}
\multiput(146.19,86.73)(-0.11,-0.16){8}{\line(0,-1){0.16}}
\multiput(145.29,85.42)(-0.13,-0.12){9}{\line(-1,0){0.13}}
\multiput(144.11,84.34)(-0.20,-0.11){7}{\line(-1,0){0.20}}
\multiput(142.72,83.55)(-0.38,-0.11){4}{\line(-1,0){0.38}}
\put(141.19,83.10){\line(-1,0){1.59}}
\multiput(139.60,83.01)(-0.52,0.09){3}{\line(-1,0){0.52}}
\multiput(138.03,83.28)(-0.25,0.10){6}{\line(-1,0){0.25}}
\multiput(136.56,83.90)(-0.16,0.12){8}{\line(-1,0){0.16}}
\multiput(135.27,84.84)(-0.12,0.13){9}{\line(0,1){0.13}}
\multiput(134.22,86.05)(-0.11,0.20){7}{\line(0,1){0.20}}
\multiput(133.48,87.46)(-0.10,0.39){4}{\line(0,1){0.39}}
\put(133.07,89.00){\line(0,1){1.60}}
\multiput(133.03,90.60)(0.11,0.52){3}{\line(0,1){0.52}}
\multiput(133.34,92.16)(0.11,0.24){6}{\line(0,1){0.24}}
\multiput(134.01,93.61)(0.11,0.14){9}{\line(0,1){0.14}}
\multiput(134.98,94.88)(0.14,0.11){9}{\line(1,0){0.14}}
\multiput(136.22,95.89)(0.24,0.12){6}{\line(1,0){0.24}}
\multiput(137.65,96.59)(0.59,0.10){4}{\line(1,0){0.59}}
\multiput(9.00,91.00)(0.79,-0.09){2}{\line(1,0){0.79}}
\multiput(10.59,90.82)(0.30,-0.11){5}{\line(1,0){0.30}}
\multiput(12.09,90.28)(0.17,-0.11){8}{\line(1,0){0.17}}
\multiput(13.43,89.42)(0.11,-0.11){10}{\line(0,-1){0.11}}
\multiput(14.54,88.27)(0.12,-0.20){7}{\line(0,-1){0.20}}
\multiput(15.37,86.91)(0.10,-0.30){5}{\line(0,-1){0.30}}
\multiput(15.86,85.39)(0.07,-0.80){2}{\line(0,-1){0.80}}
\multiput(16.00,83.80)(-0.11,-0.79){2}{\line(0,-1){0.79}}
\multiput(15.77,82.22)(-0.12,-0.30){5}{\line(0,-1){0.30}}
\multiput(15.19,80.73)(-0.11,-0.16){8}{\line(0,-1){0.16}}
\multiput(14.29,79.42)(-0.13,-0.12){9}{\line(-1,0){0.13}}
\multiput(13.11,78.34)(-0.20,-0.11){7}{\line(-1,0){0.20}}
\multiput(11.72,77.55)(-0.38,-0.11){4}{\line(-1,0){0.38}}
\put(10.19,77.10){\line(-1,0){1.59}}
\multiput(8.60,77.01)(-0.52,0.09){3}{\line(-1,0){0.52}}
\multiput(7.03,77.28)(-0.25,0.10){6}{\line(-1,0){0.25}}
\multiput(5.56,77.90)(-0.16,0.12){8}{\line(-1,0){0.16}}
\multiput(4.27,78.84)(-0.12,0.13){9}{\line(0,1){0.13}}
\multiput(3.22,80.05)(-0.11,0.20){7}{\line(0,1){0.20}}
\multiput(2.48,81.46)(-0.10,0.39){4}{\line(0,1){0.39}}
\put(2.07,83.00){\line(0,1){1.60}}
\multiput(2.03,84.60)(0.11,0.52){3}{\line(0,1){0.52}}
\multiput(2.34,86.16)(0.11,0.24){6}{\line(0,1){0.24}}
\multiput(3.01,87.61)(0.11,0.14){9}{\line(0,1){0.14}}
\multiput(3.98,88.88)(0.14,0.11){9}{\line(1,0){0.14}}
\multiput(5.22,89.89)(0.24,0.12){6}{\line(1,0){0.24}}
\multiput(6.65,90.59)(0.59,0.10){4}{\line(1,0){0.59}}
\multiput(49.00,91.00)(0.79,-0.09){2}{\line(1,0){0.79}}
\multiput(50.59,90.82)(0.30,-0.11){5}{\line(1,0){0.30}}
\multiput(52.09,90.28)(0.17,-0.11){8}{\line(1,0){0.17}}
\multiput(53.43,89.42)(0.11,-0.11){10}{\line(0,-1){0.11}}
\multiput(54.54,88.27)(0.12,-0.20){7}{\line(0,-1){0.20}}
\multiput(55.37,86.91)(0.10,-0.30){5}{\line(0,-1){0.30}}
\multiput(55.86,85.39)(0.07,-0.80){2}{\line(0,-1){0.80}}
\multiput(56.00,83.80)(-0.11,-0.79){2}{\line(0,-1){0.79}}
\multiput(55.77,82.22)(-0.12,-0.30){5}{\line(0,-1){0.30}}
\multiput(55.19,80.73)(-0.11,-0.16){8}{\line(0,-1){0.16}}
\multiput(54.29,79.42)(-0.13,-0.12){9}{\line(-1,0){0.13}}
\multiput(53.11,78.34)(-0.20,-0.11){7}{\line(-1,0){0.20}}
\multiput(51.72,77.55)(-0.38,-0.11){4}{\line(-1,0){0.38}}
\put(50.19,77.10){\line(-1,0){1.59}}
\multiput(48.60,77.01)(-0.52,0.09){3}{\line(-1,0){0.52}}
\multiput(47.03,77.28)(-0.25,0.10){6}{\line(-1,0){0.25}}
\multiput(45.56,77.90)(-0.16,0.12){8}{\line(-1,0){0.16}}
\multiput(44.27,78.84)(-0.12,0.13){9}{\line(0,1){0.13}}
\multiput(43.22,80.05)(-0.11,0.20){7}{\line(0,1){0.20}}
\multiput(42.48,81.46)(-0.10,0.39){4}{\line(0,1){0.39}}
\put(42.07,83.00){\line(0,1){1.60}}
\multiput(42.03,84.60)(0.11,0.52){3}{\line(0,1){0.52}}
\multiput(42.34,86.16)(0.11,0.24){6}{\line(0,1){0.24}}
\multiput(43.01,87.61)(0.11,0.14){9}{\line(0,1){0.14}}
\multiput(43.98,88.88)(0.14,0.11){9}{\line(1,0){0.14}}
\multiput(45.22,89.89)(0.24,0.12){6}{\line(1,0){0.24}}
\multiput(46.65,90.59)(0.59,0.10){4}{\line(1,0){0.59}}
\multiput(29.00,106.00)(0.79,-0.09){2}{\line(1,0){0.79}}
\multiput(30.59,105.82)(0.30,-0.11){5}{\line(1,0){0.30}}
\multiput(32.09,105.28)(0.17,-0.11){8}{\line(1,0){0.17}}
\multiput(33.43,104.42)(0.11,-0.11){10}{\line(0,-1){0.11}}
\multiput(34.54,103.27)(0.12,-0.20){7}{\line(0,-1){0.20}}
\multiput(35.37,101.91)(0.10,-0.30){5}{\line(0,-1){0.30}}
\multiput(35.86,100.39)(0.07,-0.80){2}{\line(0,-1){0.80}}
\multiput(36.00,98.80)(-0.11,-0.79){2}{\line(0,-1){0.79}}
\multiput(35.77,97.22)(-0.12,-0.30){5}{\line(0,-1){0.30}}
\multiput(35.19,95.73)(-0.11,-0.16){8}{\line(0,-1){0.16}}
\multiput(34.29,94.42)(-0.13,-0.12){9}{\line(-1,0){0.13}}
\multiput(33.11,93.34)(-0.20,-0.11){7}{\line(-1,0){0.20}}
\multiput(31.72,92.55)(-0.38,-0.11){4}{\line(-1,0){0.38}}
\put(30.19,92.10){\line(-1,0){1.59}}
\multiput(28.60,92.01)(-0.52,0.09){3}{\line(-1,0){0.52}}
\multiput(27.03,92.28)(-0.25,0.10){6}{\line(-1,0){0.25}}
\multiput(25.56,92.90)(-0.16,0.12){8}{\line(-1,0){0.16}}
\multiput(24.27,93.84)(-0.12,0.13){9}{\line(0,1){0.13}}
\multiput(23.22,95.05)(-0.11,0.20){7}{\line(0,1){0.20}}
\multiput(22.48,96.46)(-0.10,0.39){4}{\line(0,1){0.39}}
\put(22.07,98.00){\line(0,1){1.60}}
\multiput(22.03,99.60)(0.11,0.52){3}{\line(0,1){0.52}}
\multiput(22.34,101.16)(0.11,0.24){6}{\line(0,1){0.24}}
\multiput(23.01,102.61)(0.11,0.14){9}{\line(0,1){0.14}}
\multiput(23.98,103.88)(0.14,0.11){9}{\line(1,0){0.14}}
\multiput(25.22,104.89)(0.24,0.12){6}{\line(1,0){0.24}}
\multiput(26.65,105.59)(0.59,0.10){4}{\line(1,0){0.59}}
\multiput(29.00,81.00)(0.79,-0.09){2}{\line(1,0){0.79}}
\multiput(30.59,80.82)(0.30,-0.11){5}{\line(1,0){0.30}}
\multiput(32.09,80.28)(0.17,-0.11){8}{\line(1,0){0.17}}
\multiput(33.43,79.42)(0.11,-0.11){10}{\line(0,-1){0.11}}
\multiput(34.54,78.27)(0.12,-0.20){7}{\line(0,-1){0.20}}
\multiput(35.37,76.91)(0.10,-0.30){5}{\line(0,-1){0.30}}
\multiput(35.86,75.39)(0.07,-0.80){2}{\line(0,-1){0.80}}
\multiput(36.00,73.80)(-0.11,-0.79){2}{\line(0,-1){0.79}}
\multiput(35.77,72.22)(-0.12,-0.30){5}{\line(0,-1){0.30}}
\multiput(35.19,70.73)(-0.11,-0.16){8}{\line(0,-1){0.16}}
\multiput(34.29,69.42)(-0.13,-0.12){9}{\line(-1,0){0.13}}
\multiput(33.11,68.34)(-0.20,-0.11){7}{\line(-1,0){0.20}}
\multiput(31.72,67.55)(-0.38,-0.11){4}{\line(-1,0){0.38}}
\put(30.19,67.10){\line(-1,0){1.59}}
\multiput(28.60,67.01)(-0.52,0.09){3}{\line(-1,0){0.52}}
\multiput(27.03,67.28)(-0.25,0.10){6}{\line(-1,0){0.25}}
\multiput(25.56,67.90)(-0.16,0.12){8}{\line(-1,0){0.16}}
\multiput(24.27,68.84)(-0.12,0.13){9}{\line(0,1){0.13}}
\multiput(23.22,70.05)(-0.11,0.20){7}{\line(0,1){0.20}}
\multiput(22.48,71.46)(-0.10,0.39){4}{\line(0,1){0.39}}
\put(22.07,73.00){\line(0,1){1.60}}
\multiput(22.03,74.60)(0.11,0.52){3}{\line(0,1){0.52}}
\multiput(22.34,76.16)(0.11,0.24){6}{\line(0,1){0.24}}
\multiput(23.01,77.61)(0.11,0.14){9}{\line(0,1){0.14}}
\multiput(23.98,78.88)(0.14,0.11){9}{\line(1,0){0.14}}
\multiput(25.22,79.89)(0.24,0.12){6}{\line(1,0){0.24}}
\multiput(26.65,80.59)(0.59,0.10){4}{\line(1,0){0.59}}
\put(9.00,84.00){\makebox(0,0)[cc]{$M_0$}}
\put(49.00,84.00){\makebox(0,0)[cc]{$M_{12}$}}
\put(29.00,99.00){\makebox(0,0)[cc]{$M_2$}}
\put(29.00,74.00){\makebox(0,0)[cc]{$M_1$}}
\put(42.00,73.39){\makebox(0,0)[cc]{$\teta$}}
\put(17.00,73.39){\makebox(0,0)[cc]{$R$}}
\put(16.33,95.79){\makebox(0,0)[cc]{$\varphi$}}
\put(40.00,95.79){\makebox(0,0)[cc]{$\psi$}}
\put(70.00,70.00){\makebox(0,0)[cc]{$M_0$}}
\put(120.00,70.00){\makebox(0,0)[cc]{$M_1$}}
\put(90.00,90.00){\makebox(0,0)[cc]{$M_2$}}
\put(140.00,90.00){\makebox(0,0)[cc]{$M_{12}$}}
\put(134.67,76.79){\makebox(0,0)[cc]{$R_1'$}}
\put(111.00,96.79){\makebox(0,0)[cc]{$R_2'$}}
\put(77.33,87.19){\makebox(0,0)[cc]{$R_2$}}
\put(95.00,63.19){\makebox(0,0)[cc]{$R_1$}}
\put(64.33,95.19){\makebox(0,0)[cc]{$\teta$}}
\put(85.00,110.00){\makebox(0,0)[cc]{$\teta_2$}}
\put(124.67,100.79){\makebox(0,0)[cc]{$\teta_1$}}
\put(146.00,115.19){\makebox(0,0)[cc]{$\teta'$}}
\put(105.00,45.00){\makebox(0,0)[cc]{Fig. 5}}
\put(29.00,44.00){\makebox(0,0)[cc]{Fig. 4}}
\put(23.67,94.19){\line(-5,-3){9.67}}
\put(15.33,81.39){\line(2,-1){7.67}}
\put(34.67,77.39){\line(5,2){8.00}}
\put(42.67,87.79){\line(-4,3){8.33}}
\put(70.00,113.59){\line(0,-1){36.80}}
\put(77.00,70.39){\line(1,0){36.00}}
\put(133.00,90.39){\line(-1,0){35.67}}
\put(90.00,96.79){\line(0,1){36.80}}
\put(97.00,139.99){\line(1,0){36.00}}
\put(140.00,133.59){\line(0,-1){36.80}}
\put(120.00,76.79){\line(0,1){36.80}}
\put(113.00,119.99){\line(-1,0){36.00}}
\put(135.00,135.00){\line(-1,-1){10.67}}
\put(125.33,75.19){\line(1,1){10.33}}
\put(85.33,84.79){\line(-6,-5){10.67}}
\put(85.33,134.39){\line(-5,-4){11.00}}
\put(36.67,120.67){\makebox(0,0)[cc]{$\psi=-u \teta/\varphi$}}
\put(70.33,120.00){\makebox(0,0)[cc]{$M_3$}}
\end{picture}

This is illustrated on Fig. 5 where one can easily tag the
untagged connecting lines following the expression for
 $\psi$ on Fig. 4, commutativity of the diagrams on Figs. 4, 5
is easily verified. So the cubic diagram on Figs. 2, 5 provide us with
the method of
exclusion of  quadratures in the superposition formulas for the
cases of \R\ and Moutard \tr s. One may suppose that this diagram
gives some algebraic superposition formulas also for \B\ \tr s of
other  $(2+1)$-dimensional integrable systems (provided they
possess the "traditional" superposition, Fig. 1). Thus the Table
1 shall be completed with the following row (cf.
 \cite{ts-dis} about the validity of the  column corresponding
 to $(1+1)$-dimensional systems):

\begin{table}[h]
\begin{tabular}{|c|p{3.5cm}|p{5.5cm}|p{5.5cm}|}
\hline
4   & Solution $x^{(123)}$ on Fig. 2   & exists and is
algebraically expressible through the other given solutions
 & exists and is
algebraically expressible through the other given solutions \\ \hline
\end{tabular}
\label{tab2}
\end{table}

As one can  check using (\ref{KVADR3}), if we have a forth
initial \R\ \tr\
$\vec x^{(0)} \sar{\varphi^{(4)}}\vec x^{(4)}$
we obtain a commutative diagram of \R\ \tr s (4-dimensional cube    skeleton)
comprising 16 vertices $\vec x^{(i)}$, $\vec x^{(ij)}$,
$\vec x^{(ijk)}$, $\vec x^{(1234)}$, $1 \leq i < j<k \leq 4$.
One may conjecture validity of analogous commutative diagrams
for the case of $n$ initial \R\ \tr s
$\vec x^{(0)} \sar{\varphi^{(i)}}\vec x^{(i)}$, $ 1 \leq i \leq n$.
The last (but not the least) property of such "B\"acklund hypercube"
formulas consists in the possibility to obtain wide classes
of solutions of $(2+1)$-dimensional integrable systems in
question using only algebraic formulas (and performing quadratures
only on "the first level" which is usually trivial for a trivial
initial seed solution $u^{(0)}$).
As we have shown elsewhere (see \cite{ganzha1,ganzha2}),
one can obtain in such a way "almost all" their solutions.

\end{document}